\documentclass[preprint,12pt]{amsart}
\usepackage{epsfig,graphicx,subfigure,epstopdf,color}
\usepackage{amsmath, amssymb,mathabx,mathrsfs,amsthm}
\usepackage{hyperref}
\usepackage[toc,page]{appendix}
\usepackage{comment}
\usepackage{csvsimple}
\usepackage{float}

\newtheorem{thm}{Theorem}[section]

\theoremstyle{definition}

\theoremstyle{remark}

\numberwithin{equation}{section}

\newtheorem{tab}[thm]{Table}
\theoremstyle{definition}

\newcommand{\eps}{\varepsilon}

\def\N{{\mathbb N}}

\def\R{{\mathbb R}}

\begin{document}

\title[Topological recognition of critical transitions]{Topological recognition of critical transitions in time series of cryptocurrencies}

\author[M. Gidea]{Marian  Gidea}
\email{Marian.Gidea@yu.edu}
\address{Yeshiva University, Department of Mathematical Sciences, New York, NY 10016, USA}

\author[D. Goldsmith]{Daniel Goldsmith}
\email{daniel.goldsmith@mail.yu.edu}
\address{Yeshiva University, Department of Mathematical Sciences, New York, NY 10033, USA}

\author[Y. Katz]{Yuri Katz}
\email{yuri.katz@spglobal.com}
\address{S\&P Global Market Intelligence, 55 Water Str., New York, NY 10040, USA}

\author[P. Roldan]{Pablo Roldan}
\email{pablo.roldan@yu.edu}
\address{Yeshiva University, Department of Mathematical Sciences, New York, NY 10033, USA}

\author[Y. Shmalo]{Yonah Shmalo}
\address{Yeshiva University, Department of Mathematical Sciences, New York, NY 10033, USA}
\email{yshmalo1@mail.yu.edu}

\begin{abstract}
We analyze the time series of four major cryptocurrencies (Bitcoin, Ethereum, Litecoin, and Ripple) before the digital market crash at the end of 2017 - beginning 2018. We introduce a methodology that combines topological data analysis with a machine learning technique -- $k$-means clustering --
in order to automatically recognize the emerging chaotic regime  in a complex system approaching a critical transition.
We first test our methodology on the complex system dynamics of a Lorenz-type attractor, and then we apply it to the four major cryptocurrencies. We find early warning signals for critical transitions in the cryptocurrency markets, even though  the relevant time series exhibit a  highly erratic behavior.
\end{abstract}

\maketitle

\section{Introduction}
Critical transitions -- abrupt shifts in the state of a system that are triggered by small perturbations -- have been observed in many complex systems, including climate, ecosystems, and financial markets; see, e.g., \cite{Scheffer2009,Scheffer2012}, and the references therein. In the case of deterministic systems, critical transitions often appear in the form of catastrophic bifurcations (see, e.g., \cite{Livnia2007,Dakos2012,Kozlowska2016})), while in noisy systems, they are associated with drastic changes in the distribution of a system's states (see, e.g., \cite{Guttal2008}). Generally, it is quite challenging to identify an approaching critical transition from real-life data, because the state of the system may show little change, and only for a short time-interval, relative to the period of observation, before reaching a critical threshold.

Detection of early signals of an approaching critical transition is particularly difficult in financial time series, where the data is intrinsically very noisy and often exhibits non-stationarity (see, e.g. \cite{Livan,Munix2012}). Moreover, since in the financial markets critical transitions can occur with only little warning, it is important to extract information from short-time windows with less than $100$ data-points.
A fundamental question is how to extract useful information from such empirical data, which will lead to the construction  of an `early warning system' for financial markets.

The geometric method of topological data analysis (TDA), which lies at the core of this paper, is free from any statistical assumptions, and is able to detect critical transitions in complex systems (see, e.g., \cite{BerwaldGidea14,BerwaldGideaVejdemo14,GideaKatz18}).
The input of the method is a point cloud -- a snapshot image of the data -- to which a geometric `shape' is associated, and topological information on that shape is extracted. The output of the method is a \emph{persistence diagram}, or a \emph{persistence landscape}, which represents a summary of all topological features displayed by the data, ranked by the `resolution level'. The method is robust, that is, small noisy perturbations added to the data result  in   small changes in the persistence landscape. For deterministic systems undergoing bifurcations, the TDA is able to recognize a change in the topology of underlying attractors.
This change is reflected by the evolving shape of the point-cloud data. The features of that shape are computable via persistent homology, and the changes in the shape can be measured via the norms of the persistence landscapes.

Using TDA to recognize different regimes of behavior of an underlining complex system has received a lot of attention in the last few years. Hence, there is a well-grounded theoretical foundation, as well as numerous experimental results; see, e.g., \cite{BerwaldGidea14,BerwaldGideaVejdemo14,Perea,Perea2015,Lum15,Munch2016,Seversky,Mischaikow-et-al,Khasawneh2018}.
However, the application of TDA to financial markets is in the early stages of development; see, e.g.,  \cite{Gidea17,GideaKatz18,Truong}.

In this paper, we apply the TDA-based method to investigate the behavior of the most capitalized  cryptocurrencies -- Bitcoin, Ethereum, Litecoin, and Ripple -- prior to the crash that occurred at the end of 2017 - early 2018. The cryptocurrencies market capitalization exploded from around \$19 billion in February 2017 to around \$800 billion in December 2017, with more than 1000 cryptocurrencies currently on the market. Despite the recent downturn, which wiped out over \$550 billion of value, it is expected that the
total market valuation for cryptocurrencies will hit \$1 trillion during 2018 (see, e.g., \cite{Smartereum}).

The trading activity of cryptocurrencies appears to be heavily influenced by popularity driven by news and social media, speculative bubbles, fraud, regulations, and policy interventions (see, e.g., \cite{Lu}). This is reflected by  wild statistical properties of the corresponding time series:
lack of stationarity, strong-term memory, leverage effect, and volatility clustering have been observed;
see the related works mentioned below. Such  features makes cryptocurrencies an ideal test-case for application of TDA, which is free from  statistical assumptions.

The pipeline of our approach is comprised of the following steps:
\begin{itemize}\item we use the time-delay coordinate embedding to obtain a multi-dimensional representation of the time-series; \item we apply a sliding window to scan the obtained multi-dimensional time series; \item for each sliding window we associate a point cloud and apply a simplicial complex filtration; \item we extract topological information from this filtration in the form of a real-valued function -- a persistence landscape; \item for each window we compute the norm of the persistence landscape;
\item   we compare the derived time series of the norms of persistence landscapes with the time series of the observed log-returns to identify patterns indicative of critical transitions.
\end{itemize}

Our TDA-based approach shows that, prior to a critical transition, the norms of the persistence landscapes undergo significant changes, and such changes can be detected even with short sliding windows. When the value of an asset undergoes a critical transition, that is, an abrupt change from one regime to a significantly different one,
there is  a significant change in the shape of the data, which can be recognized by the TDA-based method.

To capture the relation between the time series of prices/returns of each asset and the $L^1$-norms  of persistence landscapes prior to the crash, we use an unsupervised machine learning technique utilizing non-parametric, geometry based $k$-means clustering. This method was selected because, similar to TDA, it also requires no statistical assumptions of the data and is based partially on geometry. The $k$-means clustering applied to the data consisting of the normalized log-prices of each asset, the log-returns, and the $L^1$-norms of the persistence landscapes, yields an automatic identification of topologically distinct regimes emerging before the crash of each asset.

We now mention some related works. A methodology to detect bubbles in the price dynamics of cryptocurrencies (specifically, Bitcoin), using multi-scale analysis, as well as $k$-means clustering, is presented in \cite{GerlachSornette}. In \cite{Catania2018}  the predictability of cryptocurrencies time series is investigated through several alternative univariate and multivariate models in point and density forecasting. A general equilibrium monetary model of a cryptocurrency market, which endogenizes the value of a cryptocurrency, and the underlying trading and mining activities, is developed in \cite{Chiu2017}. A dynamic model of price formation is proposed in \cite{Catania2017}. Statistical properties of the Bitcoin time series are studied in \cite{Drozdz}. Statistical properties of the cryptocurrencies are explored in \cite{Chan2017}, where a wide range of parametric distributions are fitted to data. In \cite{Chu2017}, twelve GARCH models are fitted to each of the seven most popular cryptocurrencies.

The structure of the paper is as follows.
In Section \ref{sec:method}, we provide background on the methodology.
In Section \ref{sec:testing}, we test this methodology on a chaotic time series generated by  a dynamical system  (with small additive noise) undergoing bifurcations. Our experiments show that that TDA is able to detect critical transitions in the time series, as the $L^1$-norms follow an increasing trend when the system transitions from a less turbulent to a more turbulent regime. Additional examples of this behavior can be found in \cite{GideaKatz18,BerwaldGidea14,BerwaldGideaVejdemo14}. Since these time series are simulated, it is easy to match the changes in the norms with the changes in the parameters driving the system. Topological recognition of critical transitions becomes much more difficult in real-life financial time series with no precise demarcation lines between different regimes. For this reason, we  apply a $k$-means clustering technique to recognize critical transitions.
In Section \ref{sec:cryptocurrencies}, we apply our methodology to the  time series of the log-returns of Bitcoin, Ethereum, Litecoin, and Ripple between January 2016 and January 2018, and focus on recognition of different topological patterns associated to the critical transition prior to the crash. The $k$-means clustering method is then applied to a data set consisting of $3$ variables: the normalized value of the price of each asset, the log-return of asset, and $L^1$-norm of the persistence landscape. The outcome of this automatic classifier is the identification of clusters of topologically distinct regimes, which correspond to relevant time periods prior to the crash.
In Section \ref{section:conclusion}, we conclude that the proposed method is able to recognize critical transitions in time-series of the cryptocurrencies under consideration, and can be used to identify specific time-periods associated to such transitions.

The highlights of our work  are:
\begin{itemize}
\item	Our approach can be applied  to time series generated by complex systems that exhibit strong non-linearity and non-stationarity;
\item The topological tools involved are robust under small additive noise;
\item	The method is able to recognize critical transitions even when the utilized sliding windows are short.
\end{itemize}

\section{Methodology}\label{sec:method}
\subsection{Time-delay coordinate embedding}\label{sec:takens}
Time-delay coordinate embedding is used to obtain a  description of  the phase space of  a nonlinear dynamical system from the time series of  certain observables of the system; see, e.g., \cite{Packard,Kennel,Muldoon}

We briefly review the method. Given a discrete dynamical system defined by a diffeomorphism $f:M\to M$   on a $D$-dimensional manifold $M$, the evolution of a state $p\in M$ is given by  its orbit  $\{f^t(p)\,:\,t\in\mathbb{Z}\}$.
For an observable $\phi:M\to\mathbb{R}$ of the system,    one defines a reconstructed set $\mathscr{A}^d$ consisting of $d$-dimensional vectors of the form
\[(\phi(p), \phi(f (p)), \ldots , \phi(f^{d-2}(p)),\phi(f^{d-1}(p)))\] which are
obtained by successive evaluations of $\phi$ along orbit segments
\[\{p, f(p), \ldots ,  f^{d-2}(p), f^{d-1}(p)  \}.\]
Assume that $f$, $\phi$ are $C^r$-differentiable with $r\geq 2$. The classical theorem of Takens \cite{Takens1981} says that, if $d\geq 2D+1$, then for generic ($C^1$-open and dense) sets of $f$ and $\phi$, the map
\[p\in M \mapsto \Phi(p):=(\phi(p), \phi(f (p)), \ldots , \phi(f^{d-2}(p)),\phi(f^{d-1}(p)))\]
is an embedding.    Moreover, the map $f$ on $M$ is topologically conjugate with the left-shift map $\sigma$ on the reconstructed set $\mathscr{A}^d$, which is defined by
\[(\phi(p),   \ldots , \phi(f^{d-2}(p)),\phi(f^{d-1}(p)))\mapsto ( \phi(f (p)), \ldots , \phi(f^{d-1}(p)),\phi(f^{d}(p))).\]
Topological conjugacy means that $\Phi$ matches orbits of $f$ with orbits of $\sigma$, i.e.,  $\sigma\circ\Phi=\Phi\circ f$.

Takens' Theorem has been extended in \cite{Sauer} to the case of an invariant set (e.g., an attractor) $A\subseteq M$,  in which case it is sufficient to consider   delay-coordinate vectors of dimension $d\geq 2D+1$,  where $D$ is the box-counting dimension of the invariant set. Also, the genericity with respect to  $f$ and $\phi$ is replaced with a concept of `prevalence'. The reconstructed set $\mathscr{A}^d$  is topologically
equivalent to $A$. An important consequence is that the topology of $A$ can be described in terms of the homology groups of $\mathscr{A}^d$.   Thus, bifurcations of  attractors can be detected via changes in the homology groups associated to the time-delay coordinate reconstructed sets.

We also note  that there  exist stochastic versions of Takens' Theorem, for  stochastically forced systems; see, e.g., \cite{Stark}.

In practice the embedding dimension $d$  is not known beforehand, and  the dimension $D$ of the invariant set cannot be estimated without first embedding the time series. There are practical methods to estimate the embedding dimension; see, e.g., \cite{Kennel}. However, in this paper we will only use a fixed, low dimensional embedding.

\subsection{State space reconstruction from time series}\label{sec:reconstruction}
Suppose that we have a dynamical  system, consisting of a deterministic part plus a stochastic part, of the form
\begin{equation}\label{eqn:system}
p_{t+1}=f(p_t;\lambda_t)+\epsilon\omega_t,\end{equation}
where the map $p\mapsto f(p;\lambda)$ represents a parameter-dependent dynamical system,  $\lambda_t$ is a deterministically time-varying parameter, $\omega_t$ is a normally distributed random variable, and $\epsilon>0$ is the size of the random perturbation.

We assume that for the deterministic  part of the dynamical system (no  noise) the system parameter $\lambda_t$  evolves slowly in time,  in a manner which we now describe.
When  the  parameter is frozen at some value $\lambda_t=\lambda$, the orbits of $p_{t+1}=f(p_t;\lambda)$ starting from any initial condition $p_0$ within some `basin of attraction'  approach an  attractor $A_{\lambda}$, e.g., an attractive fixed point, or a periodic orbit, or a chaotic attractor.
When the parameter $\lambda_t$ varies slowly in time, that is, at each time-step the parameter changes by a small increment $\Delta \lambda_t=\lambda_{t+1}-\lambda_t$,
the state $p_t$ will follow closely the `instantaneous attractor' $A_{\lambda_t}$.
Specifically, we assume that the increment $|\Delta \lambda_t|$ at each time-step
is small enough so that the corresponding `instantaneous attractors' are topologically  close to one another; more precisely,  we require that
the absolute values of the differences between the norms of the persistence landscapes (defined in Section \ref{sec:TDA}) corresponding to the `instantaneous attractors' at consecutive instants of time are less than some pre-defined smallness parameter.  Of course, all the parameters mentioned above depend on the specific dynamical system that is considered.

As the parameter  of the (no-noise) system is slowly evolving in time, the `instantaneous attractor' is also evolving  in time and is undergoing bifurcations at some parameter values, thus changing its topology.  The orbits of the system closely follow  the time-varying  `instantaneous attractor', tracking its  changes in topology along the way.

If small noise is added to the deterministic system with the slowly evolving parameter, the  orbits of the system  oscillate  around  the `instantaneous attractors',
and jump between one  `instantaneous attractor' and another when approaching a bifurcation --  thus experiencing  a critical transition.
Moreover, since at a  bifurcation  there is a loss of attractivity, the orbits   are typically experiencing  increasing  oscillations prior to the  bifurcation -- thus supplying early signals of critical transitions.
An explicit example is given in Section \ref{sec:testing}.

We now explain how to use the time series corresponding to a system as above  to track  changes in the topology of the evolving `instantaneous attractor'.
Consider a corresponding time series
\[X=\{x_0,x_1,x_2,\ldots, x_{N-1} \},\]
representing measurements $x_t=\phi(p_t)$ of a certain observable $\phi$ of the system \eqref{eqn:system} at the time instants $t=0,1,\ldots,N-1$.
We will now think of the underlying dynamical system as a `black box', i.e., it is
unknown to the observer.

We transform the time series $X$ into a sequence of $(N-d+1)$  $d$-dimensional time-delay coordinate vectors \begin{equation*}\begin{split}z_0&=(x_0,x_{1},\ldots,x_{d-1}),\\ z_1&=(x_1,x_{2},\ldots,x_{d}),\\  &\cdots \\z_{t}&= (x_{t},x_{t+1},\ldots,x_{t+d-1}),\\ &\cdots \\ z_{N-d}&=(x_{N-d},x_{N-d+1},\ldots,x_{N-1}),\end{split}\end{equation*} where the embedding dimension $d$ is suitably chosen. In our applications below, we will only be interested in detecting   $D=1$-dimensional objects (loops), in which case we can safely choose $d\geq 3$, due to Takens' theorem.

We apply a sliding window $Z^t=\{z_t,z_{t+1},\ldots, z_{t+w-1}\}$ of size $w$, for $t\in\{0,\ldots,N-d-w+1\}$, with $w$ sufficiently large with $d\ll w\ll N$. The vectors $\{z_t,z_{t+1},\ldots, z_{t+w-1}\}$  form a time-varying point-cloud embedded in  $\mathbb{R}^d$. This point cloud  typically lies in some small neighborhood of the `instantaneous attractor' $A_{\lambda_t}$, except for  $\lambda_t$'s close to the bifurcation values, where the point cloud may veer off the attractor.

The evolution in time of the system is now described by the point-cloud $Z^t$ of $w$ points in $\mathbb{R}^d$, which varies in time.   The size $w$ of the window is chosen empirically, depending  on the system considered. The heuristic criterion is that, when  $\lambda_t$ is far from the bifurcation values, the point clouds of size $w$ should be large enough so that they  capture the topology of the `instantaneous attractors'   $A_{\lambda_t}$; on the other hand, if $w$ is too large, then the point clouds will meld  topological information corresponding to  multiple `instantaneous attractors', which may vary substantially across the window.

In Section \ref{sec:TDA} we will describe a method that provides a topological representation of each time-varying point-cloud $Z^t$. The outcome of applying this method to a point cloud is a persistence diagram (which represents a point in some metric space), or a persistence landscape (which represents a point in some Banach space). We  use
this topological information to detect early signals of critical transitions.
Heuristically, when the underlying dynamical system undergoes a critical transition, there is an increase in the norms of the persistence landscapes of the corresponding point clouds.

\subsection{Topological data analysis of time-varying point-clouds}\label{sec:TDA}
In this section we  briefly review the ideas of persistence homology and  persistence landscapes applied to point-clouds. Technical details on this method  can be found in, e.g,,  \cite{Cohen-Steiner,Carlsson09,Edelsbrunner10,Bubenik15,Chazal2015,Bubenik16}.

For each instant of time $t\in\{1,\ldots,N-d-w+2\}$, we  consider the corresponding point-cloud $\{Z^t\}\subseteq \mathbb{R}^d$ from  Section \ref{sec:reconstruction}.
To simplify the notation, let us fix such a point-cloud  and denote it by $Z = \{z_0, \ldots ,z_{w-1}\}$. We associate to it a topological space as follows. Introduce a `resolution'  parameter  $\eps>0$. Define the so called
Vietoris-Rips simplicial complex $R(Z, \eps)$, or, simply Rips complex, obtained as follows:
\begin{itemize}\item For each $k= 0,1, \ldots$, a $k$-simplex of vertices $\{z_{i_0},\ldots, z_{i_{k}}\}$ is part of $R(Z,\eps)$ if and only if the mutual distance between  any pair of its   vertices is less than~$\eps$, that is \[d(z_{i_j},z_{i_l})<\eps, \textrm { for all } z_{i_j},z_{i_l}\in \{z_{i_0},\ldots, z_{i_{k}}\}. \]\end{itemize} In other words, a $k$-simplex is included in $R(Z,\eps)$   whenever the vertices of that simplex are `indistinguishable from one another' at   resolution level of $\eps$.

The Rips simplicial complexes $R(Z, \eps)$ form a filtration, that is, $R(Z,\eps)\subseteq R(Z,\eps')$ whenever $\eps<\eps'$. For each such complex, we can compute its $n$-dimensional homology $H_n(R(Z,\eps))$ with coefficients in some field. Informally, the generators of the $0$-dimensional homology group $H_0(R(Z,\eps))$  correspond to the connected components of $R(Z,\eps)$, the generators of the $1$-dimensional homology group  $H_1(R(Z,\eps))$ correspond to the  `independent loops' in $R(Z,\eps)$,  the generators of the $2$-dimensional homology group $H_2(R(Z,\eps))$ correspond to  `independent cavities' in $R(Z,\eps)$, etc.
For the rest of the paper we will use only  $1$-dimensional homology.

The filtration property of the Rips complexes induces a filtration on the corresponding homologies, that is
$H_n(R(Z,\eps))\subseteq H_n(R(Z,\eps'))$ whenever $\eps<\eps'$, for each $n$. These inclusions determine canonical homomorphisms $H_n(R(Z,\eps))\hookrightarrow H_n(R(Z,\eps'))$, for $\eps<\eps'$. Due to this family of induced mappings,  for each non-zero $n$-dimensional homology class $\alpha$  there exists a pair of  values $\eps_1<\eps_2$, such that:
\begin{itemize}\item   $\alpha \in H_n(R(Z,\eps_1))$ but is not in the image of any $H_n(R(Z,\eps_1-\delta))$ under the corresponding homomorphism, for $\delta>0$,
\item the  image of $\alpha$ in $H_n(R(Z,\eps'))$ is non-zero for all $\eps_1< \eps'<\eps_2$, but the  image of $\alpha$ in $H_n(R(Z,\eps_2))$ is zero.
\end{itemize}
In this case, one says that the class $\alpha$ is `born' at the parameter value $b_\alpha:=\eps_1$, and `dies' at the parameter value $d_\alpha=\eps_2$; the pair $(b_\alpha, d_\alpha)$ represents the `birth' and `death' indices of $\alpha$. The multiplicity
$\mu_\alpha(b_\alpha,d_\alpha)$ of the point $(b_\alpha,d_\alpha)$ equals the number of classes $\alpha$ that are born at ${b_\alpha}$ and die at ${d_\alpha}$.
This multiplicity is finite since the simplicial complex is finite.

The information on  the $n$-dimensional homology generators at all scales can be encoded in a \emph{persistence diagram} $P_n$. Such a diagram consists of:
\begin{itemize}
\item  for each $n$-dimensional homology class $\alpha$ one assigns a point $p_\alpha=p_\alpha(b_\alpha,d_\alpha)\in\mathbb{R}^2$ together with its multiplicity $\mu_\alpha=\mu_\alpha(b_\alpha,d_\alpha)$;
\item in addition, $P_n$ contains all points in the positive diagonal  of $\mathbb{R}^2$; these points represent  all trivial homology generators that are born and instantly die at every level; each point on the diagonal has infinite multiplicity.
\end{itemize}
The axes of a persistence diagram are birth indices on the horizontal axis and death indices on the vertical axis.

The space of persistence diagrams can be embedded into a Banach space, whose norm can be used to derive a metric.
One such embedding is based on \emph{persistence landscapes}, consisting of sequences of functions in the  Banach space $L^p(\N\times \R)$. For each birth-death point $(b_\alpha,d_\alpha) \in P_n$, we first define a piecewise linear function
\begin{equation}\label{eqn:landscape_1}
f_{(b_\alpha,d_\alpha)}(x)=\left\{
                          \begin{array}{ll}
                            x-b_\alpha, & \hbox{if $x\in\left(b_\alpha,\frac{b_\alpha+d_\alpha}{2}\right]$;} \\
                            -x+d_\alpha, & \hbox{if $x\in\left(\frac{b_\alpha+d_\alpha}{2},d_\alpha\right)$;} \\
                            0, & \hbox{if $x\not\in (b_\alpha, d_\alpha)$.}
                          \end{array}
                        \right.
\end{equation}

To a persistence diagram $P_n$ consisting of a finite number of off-diagonal points, we associate
a sequence of functions $\lambda=(\lambda_i)_{i\in\mathbb{N}}$, where $\lambda_i:\mathbb{R}\to[0;1]$ is given by
\begin{equation}\label{eqn:landscape_2} \lambda_i(x)=i\textrm{-max}\{f_{(b_\alpha,d_\alpha)}(x)\,|\, (b_\alpha,d_\alpha)\in P_n\}
\end{equation}
where  $i\textrm{-max}$ denotes the $i$-th largest value of a  function.
We set $\lambda_i(x) = 0$ if
the $i$-th largest value does not exist. An example of  a point cloud, the corresponding persistence diagram  and persistence landscape are shown in Fig.~\ref{fig:Lorenz_persistence.png}

Via the above embedding, the persistence landscapes form a subset of the Banach space $L^p(\N\times \R)$,
where the norm of $\lambda$ is given by:
\begin{equation}\label{eqn:landscape_3}\|\lambda\|_p=\left (\sum_{i=1}^{\infty}\|\lambda_i\|_p^p\right)^{1/p}.\end{equation}
Above, $\|\cdot\|_p$ denotes the $L^p$-norm, $p\geq 1$, i.e., $\|f\|_p=\left ( \int _\mathbb{R} |f|^p\right )^{1/p}$, where the integration is with respect to  the Lebesgue measure on $\R$.
For the rest of the paper we will use only the $L^1$ norm.

The above computation of persistence diagrams and persistence landscapes is to be carried out for every sliding window $Z^t$. The output is the time series of  the $L^1$-norm of the corresponding persistence landscapes $\{\|\lambda^t_i\|_1\}_{i\geq 0}\}$. This can be summarized by the following

\medskip

\noindent\fbox{\begin{minipage}{36.5em}
\noindent $\textbf{Pipeline: }\quad
Z^t\longrightarrow \{R(Z^t,\eps)\}_{\eps}\longrightarrow \{H_*(R(Z^t,\eps))\}_{\eps} \longrightarrow  \{\lambda^t_i\}_{i\geq 0} \longrightarrow \{\|\lambda^t\|_1\}_{i\geq 0}.$

$\quad\qquad \textrm{Point-cloud} \quad\textrm{Rips complex} \quad\quad \textrm{Homology} \quad\qquad \textrm{Landscape}
\qquad \textrm{$L^1$-norm }$
\end{minipage}}

\medskip

A remarkable property that makes persistence homology suitable to analyze noisy data is its \emph{robustness} under small perturbations. Informally, this property says that if there is a small change in the underlying  point-cloud data, then the change in the  corresponding  persistence diagram is also small.

\subsection{K-means clustering}\label{section:k-means}
To capture in a precise way  the relationship between the underlying time series  and the $L^1$-norms, we use an unsupervised machine learning technique utilizing non-parametric,  geometry based $k$-means clustering. The main benefit is that this method requires no  statistical assumptions to be satisfied beforehand, in contrast with most statistical tests.

The basic algorithm starts with $k$ initial centroids (randomly chosen), and assigns each data point to the nearest centroid; each collection of points assigned to the same centroid forms a cluster.  For each cluster, the squared error  is calculated as the sum of the squares of the Euclidean distances between the data points and the cluster centroid. Then the centroid of each cluster is re-calculated and updated, and the process of assigning the data points to clusters  is repeated. The procedure is iterated until the outcome  stabilizes, i.e., the updated centroids remain the same, and no point changes cluster. This is achieved when the cluster squared error cannot be reduced any further. Different choices of the initial (random) centroids may yield different clusters. To obtain independence of the clusters on the initial choices,  one runs a large number of random initial centroids, and then select the one yielding the lowest cluster squared error. The number $k$ of clusters needs to be pre-specified at the beginning of the algorithm. There are several basic methods --  `elbow', `silhouette' and `gap statistic' -- to determine the optimal number of clusters. For a survey and references see \cite{Jain}.

We illustrate the application of this method in Section \ref{sec:testing}, and then use it in Section \ref{sec:cryptocurrencies}. As input for the $k$-means clustering, in Section \ref{sec:testing}
we will use the values of the time series of interest and the $L^1$-norms of the persistence landscapes associated to the time series.
In Section \ref{sec:cryptocurrencies} we will use the log-returns,  the $L^1$-norms of the persistence landscapes associated to the the log-returns, and the log of the price of the asset. We include this additional time series since it represents particularly relevant information for the market behavior. The output of the $k$-means clustering consists of subsets of data for which the trends  in the time series are consistent with those in the $L^1$-norms of the persistence landscapes.

\section{Noisy chaotic time series}\label{sec:testing}
We  consider time series derived from a \emph{chaotic dynamical system} dressed with noise, on which we test the methodology described above to detect critical transitions.
The system under study is given by a  three-dimensional diffeomorphism that, for certain parameter values,
possesses Lorenz-type chaotic attractors \cite{Gonchenko}.

We stress that this is different
from the well-known system of differential equations that generates the classical Lorenz attractor. We chose  this particular model because it is a discrete-time dynamical system, and generates  a $3$-dimensional attractor with interesting topology, thereby suitable for TDA. Moreover, this is a genuine chaotic attractor, in the sense that the chaotic attractor persists for open domains in the parameter space, as it does not exhibit stability windows (as in the case of the logistic map, or H\'enon map). Furthermore, this type of attractor occurs naturally in unfoldings of several types of homoclinic bifurcations, therefore is  guaranteed to be present in models from natural applications.

\begin{figure}
\centering
\includegraphics[width=0.5\textwidth]{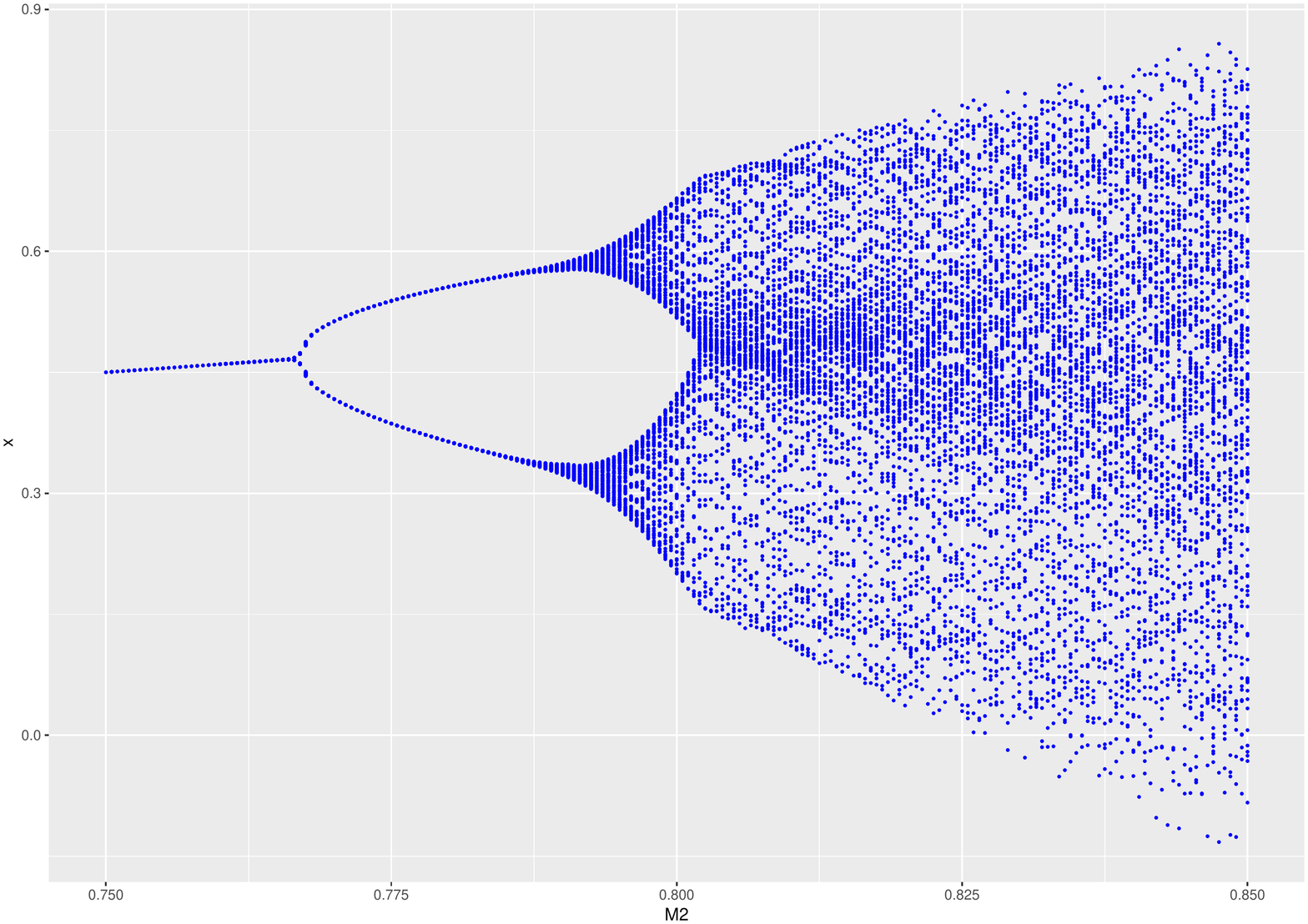}
\caption{Bifurcation diagram of the Lorenz-type map \eqref{eqn:Lorenz}, in $(M_2,x)$-coordinates. }
\label{fig:Lorenz_1}
\end{figure}

\begin{figure}\centering
$\begin{array}{cc}
\includegraphics[width=0.4\textwidth]{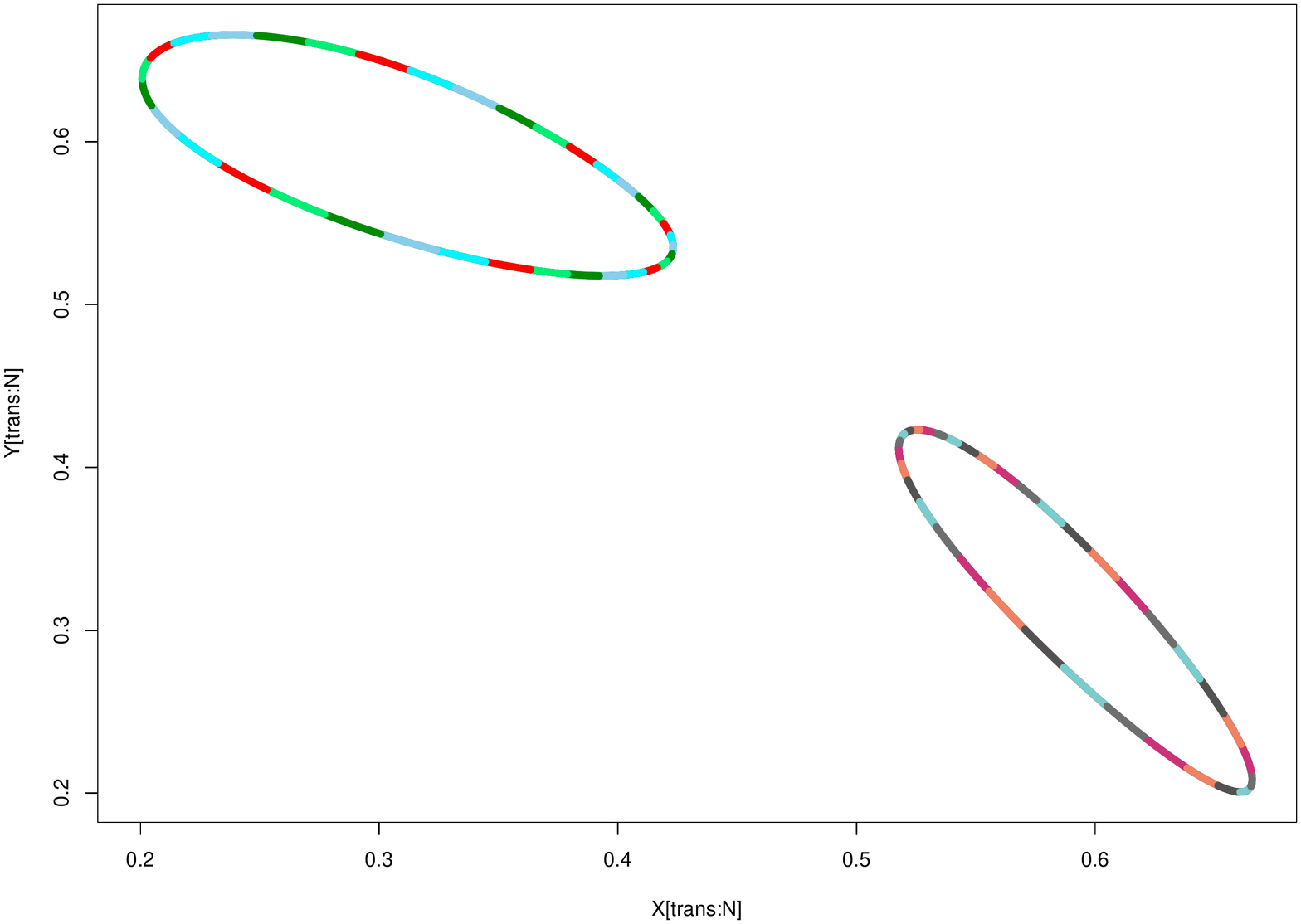}
&\includegraphics[width=0.4\textwidth, height=0.23\textheight]{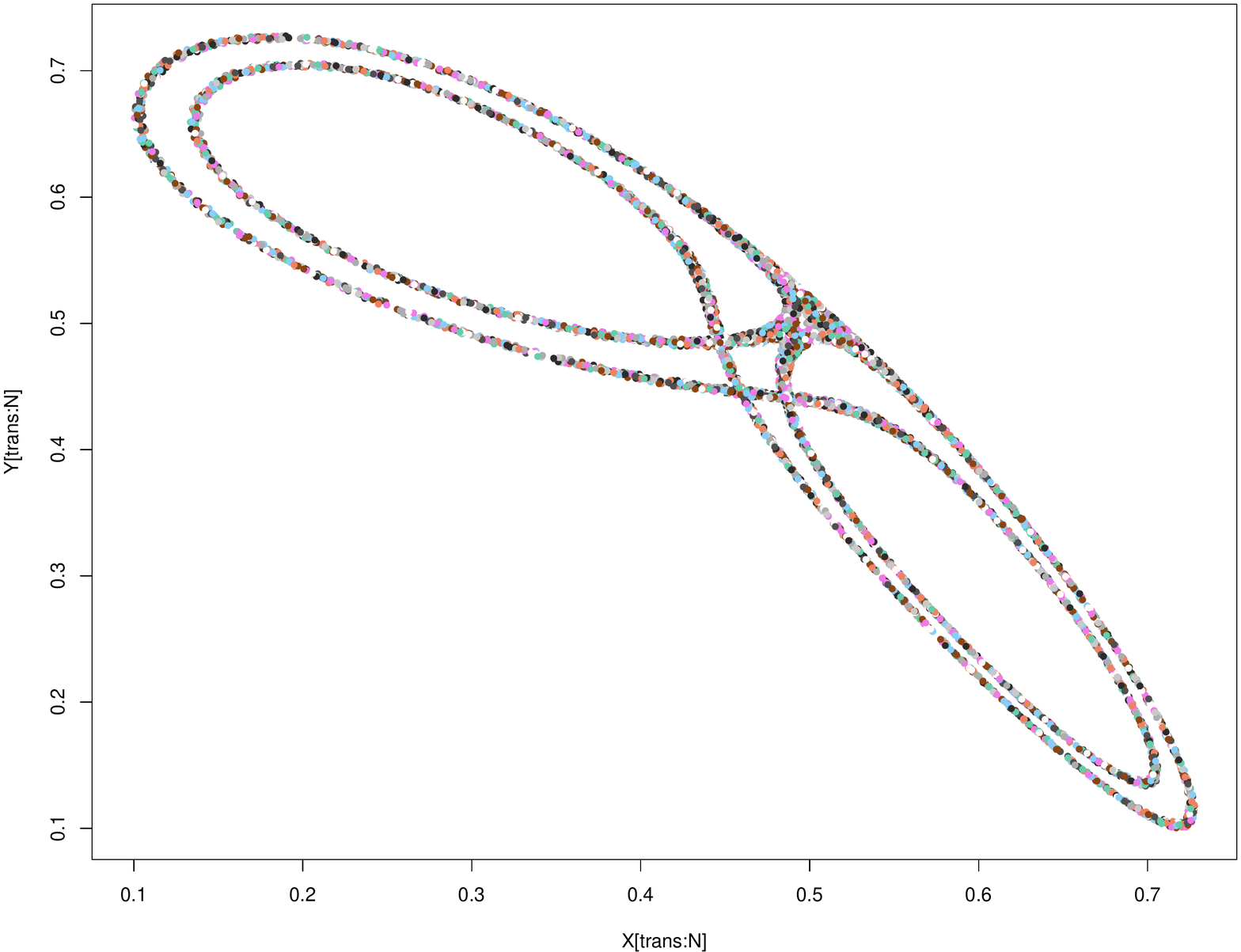}
\\
\includegraphics[width=0.4\textwidth]{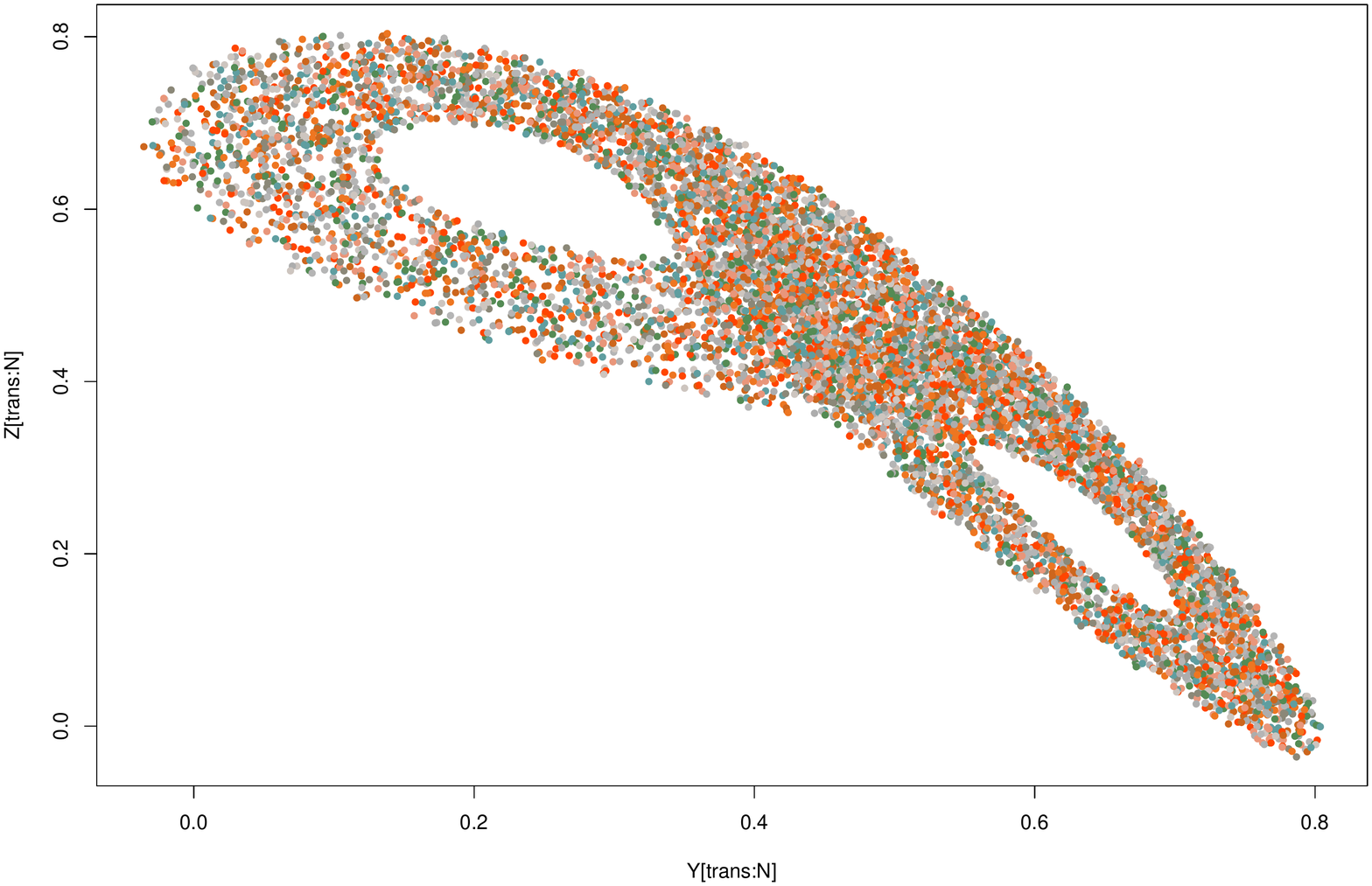}
&\includegraphics[width=0.4\textwidth]{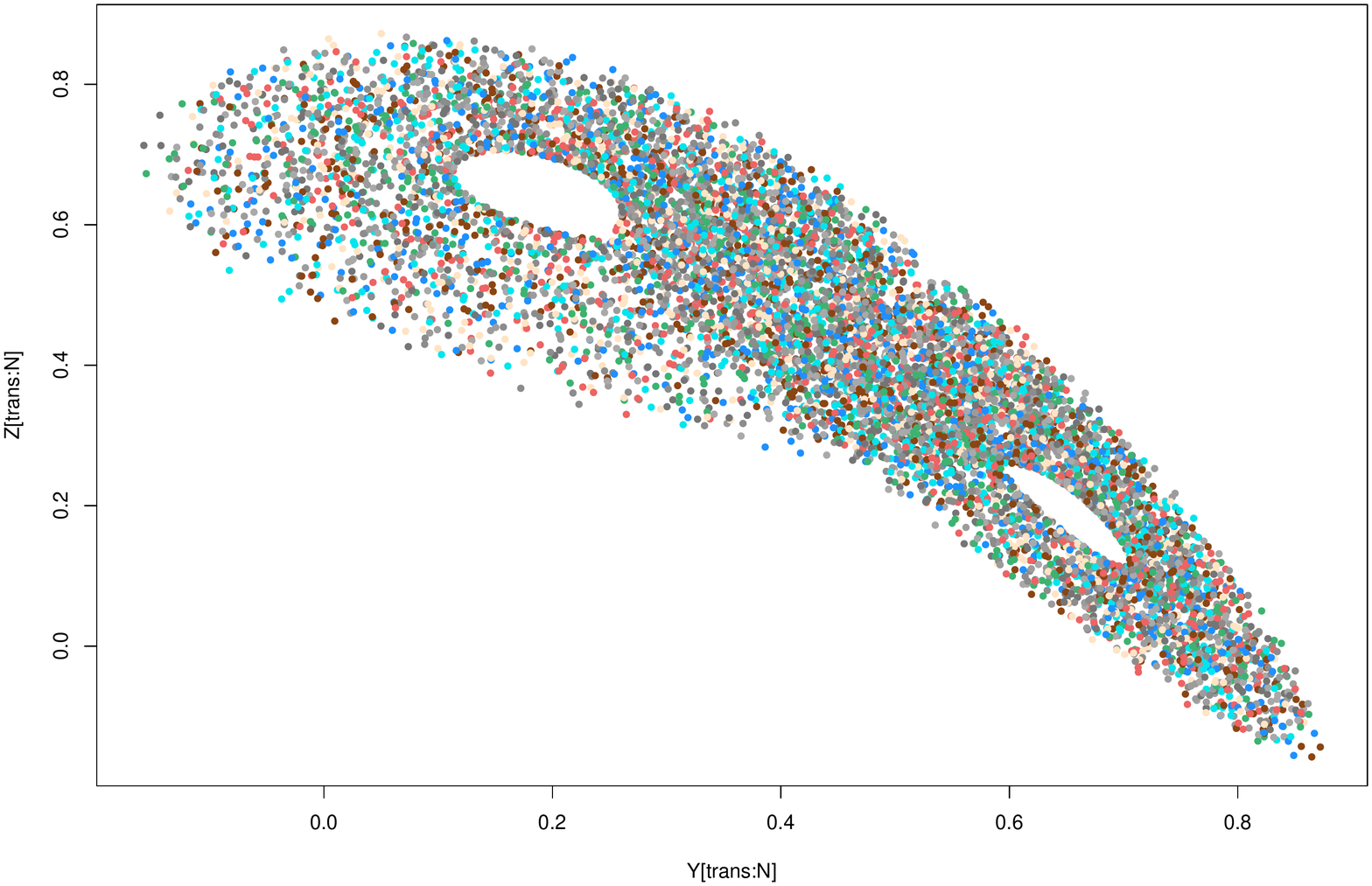}
\end{array}$
\caption{Top: Lorenz-type attractor, generated by \eqref{eqn:Lorenz},  for: (i) $M_2=0.8$ (top-left), (ii) $M_2=0.81$ (top-right), Bottom:  (iii) $M_2=0.83$ (bottom-left), (iv) $M2=0.85$ (bottom-right)}
\label{fig:Lorenz_attractor}
\end{figure}

The diffeomorphism that generates the dynamics is:
\begin{equation}\label{eqn:Lorenz}
f(x,y,z)=(y,z, M_1+Bx+M_2y-z^2),
\end{equation}
where $M_1,B,M_2$ are real parameters.
For certain values of $M_1,B,M_2$,  all suitable initial conditions approach asymptotically a chaotic attractor. When  $M_1$ and $B$ are kept fixed and $M_2$ is varied, the system   evolves from a single attractive fixed point to a pair of  attractive fixed points, then to a pair of attractive periodic orbits, and eventually ends up with a chaotic attractor.
The  bifurcation diagram for $M_1=0$, $B=0.7$, and varying $M_2$,   is shown in Fig.~\ref{fig:Lorenz_1}. Each vertical line in the bifurcation diagram represents the projection onto the $x$-coordinate  of the attractor for the corresponding $M_2$ parameter value. Several instances of the attractor  are shown in Fig.~\ref{fig:Lorenz_attractor}.

\begin{figure}\centering
$\begin{array}{cc}
\includegraphics[width=0.5\textwidth]{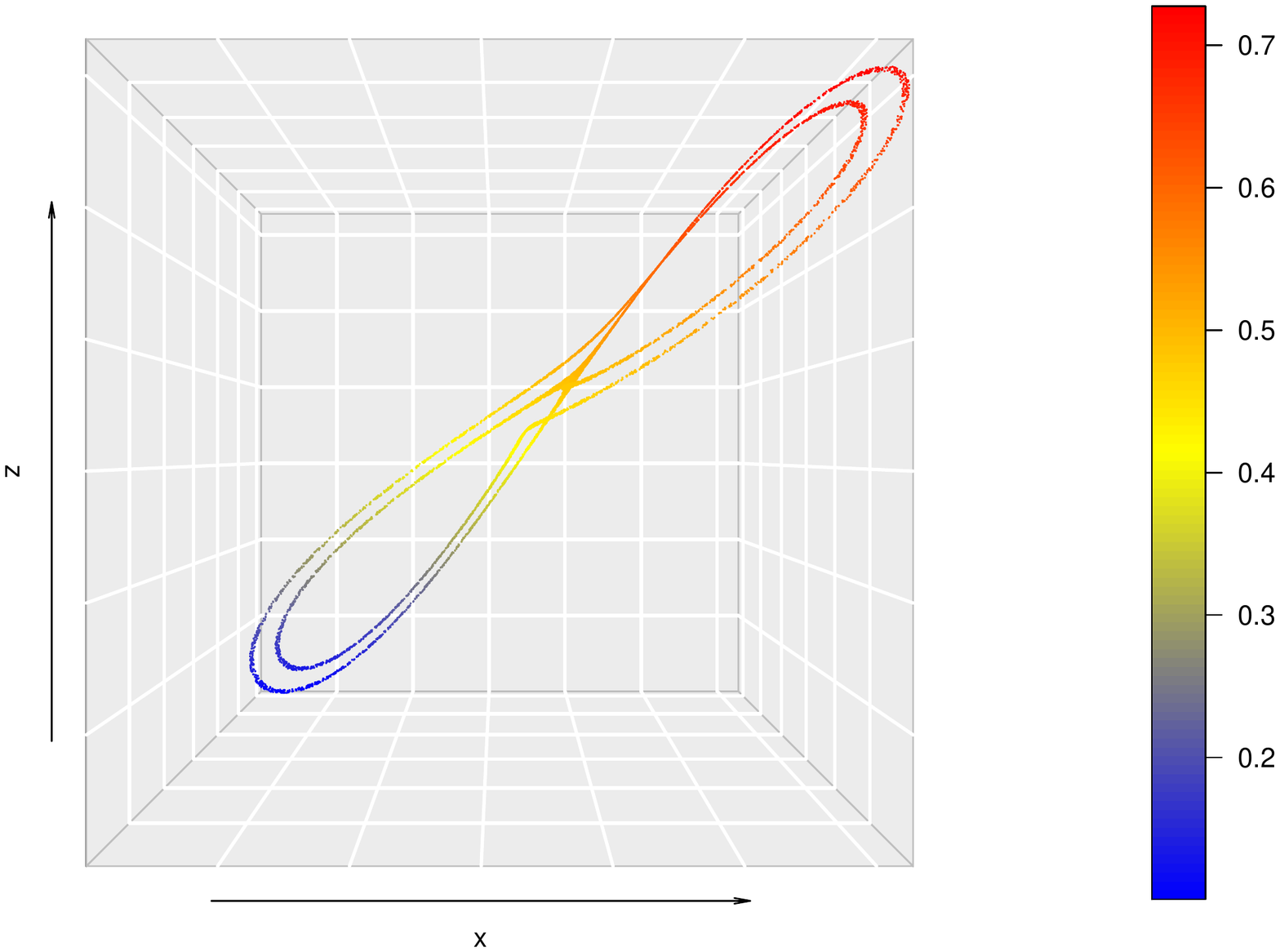}&
\includegraphics[width=0.5\textwidth]{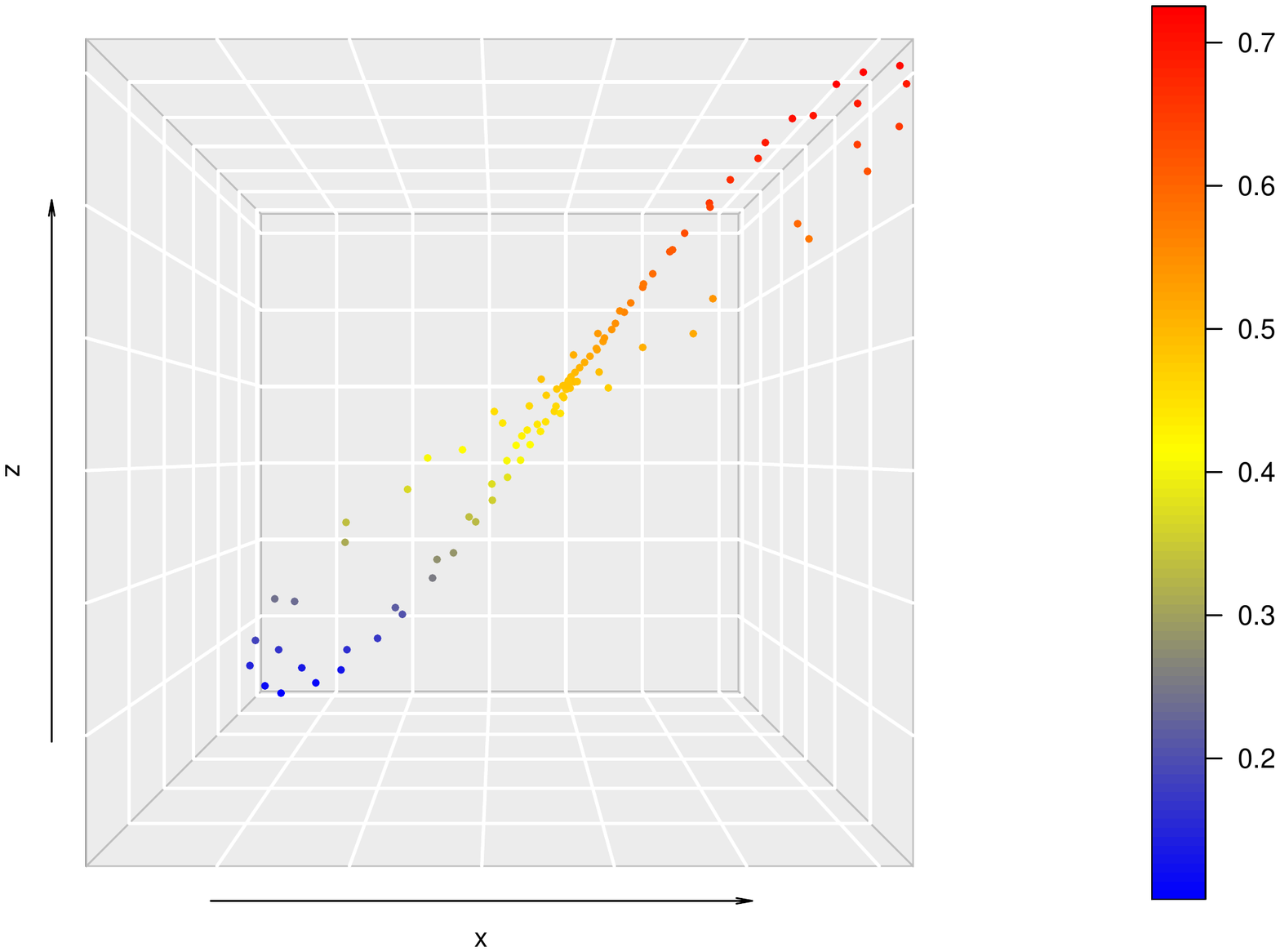}\\
\includegraphics[width=0.35\textwidth]{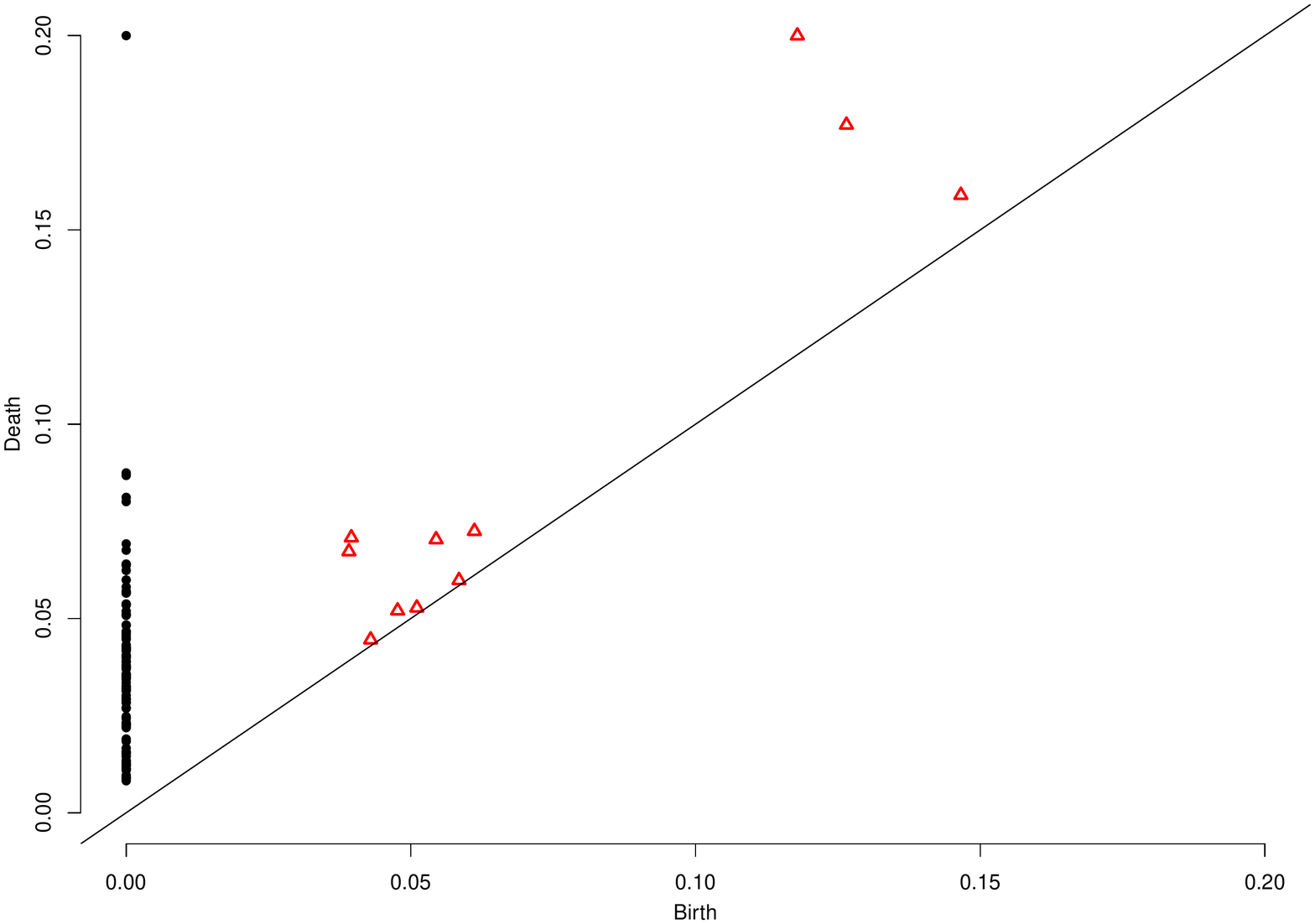}&
\includegraphics[width=0.35\textwidth]{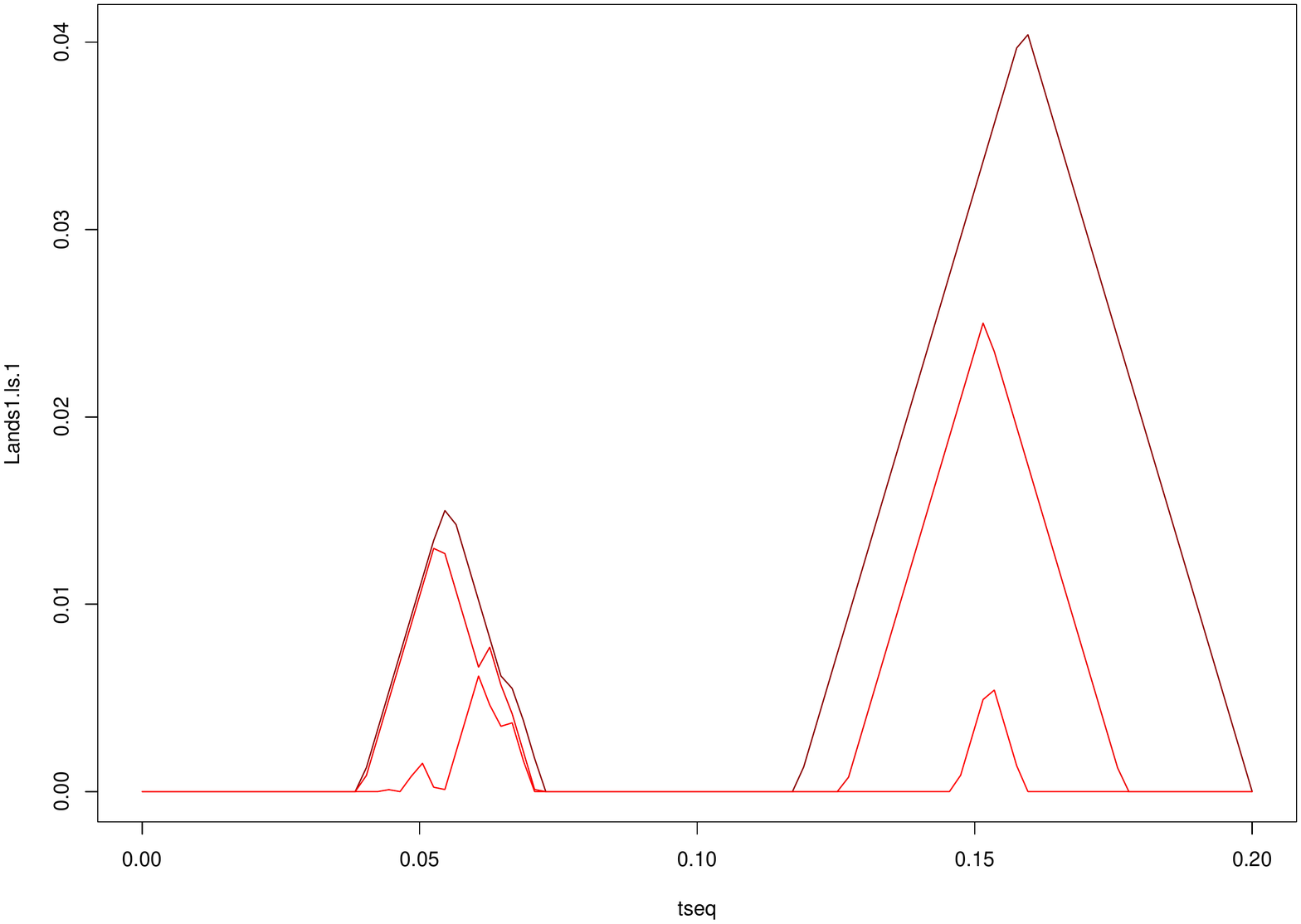}
\end{array}$
\caption{Example: (a) Lorenz-type attractor, generated by \eqref{eqn:Lorenz}, for $M_1=0$, $B=0.7$, $M_2=0.81$, (b)~reconstructed  Lorenz-type attractor, (c)~persistence diagram, where the $0$-dimensional generators are displayed as dot symbols (black), and the $1$-dimensional generators as triangle symbols (red); the two main triangle symbols farthest from the diagonal correspond to the two periodic orbits (loops); the other triangle symbols correspond to noisy loops in the point-cloud,
     (d)~$1$-dimensional persistence landscape $\lambda=(\lambda_k)_k$, for $k=1,2,3$; the two main peaks correspond to the two periodic orbits (loops), and the smaller peaks correspond to noisy loops.}
\label{fig:Lorenz_persistence.png}
\end{figure}

First, we use this model to illustrate the methodology presented in Section \ref{sec:TDA}; we generate the attractor  for $M_1=0$, $B=0.7$, $M_2=0.81$, then we reconstruct the phase space from the $x$-time series,
utilizing a 4D embedding; since we are interested only in the $1$-dimensional homology of the attractor, using $d=4$ for the embedding dimension is sufficient.
Then we compute the persistence diagram and the corresponding persistence landscape.  See Fig.~\ref{fig:Lorenz_persistence.png}.

Second,  we consider the system with a slowly evolving parameter, that is, at each step of the iteration of \eqref{eqn:Lorenz}, we increase the parameter $M_2$ by a small increment. That is:
\begin{equation}\label{eqn:Lorenz_slow}
\begin{split}
x_{t+1}=&y_t,\\
y_{t+1}=&z_t, \\
z_{t+1}=&M_1+Bx_t+M_{2,t}y_t-z_t^2,\\ M_{2,t+1}=&M_{2,t}+\Delta M_2.
\end{split}
\end{equation}
where we chose $\Delta M_2=2.8\times 10^{-5}$. We also add to the system  small Gaussian noise as in \eqref{eqn:system}; we choose the noise intensity $\epsilon=10^{-3}$.

\begin{figure}
\centering
\includegraphics[width=0.5\textwidth]{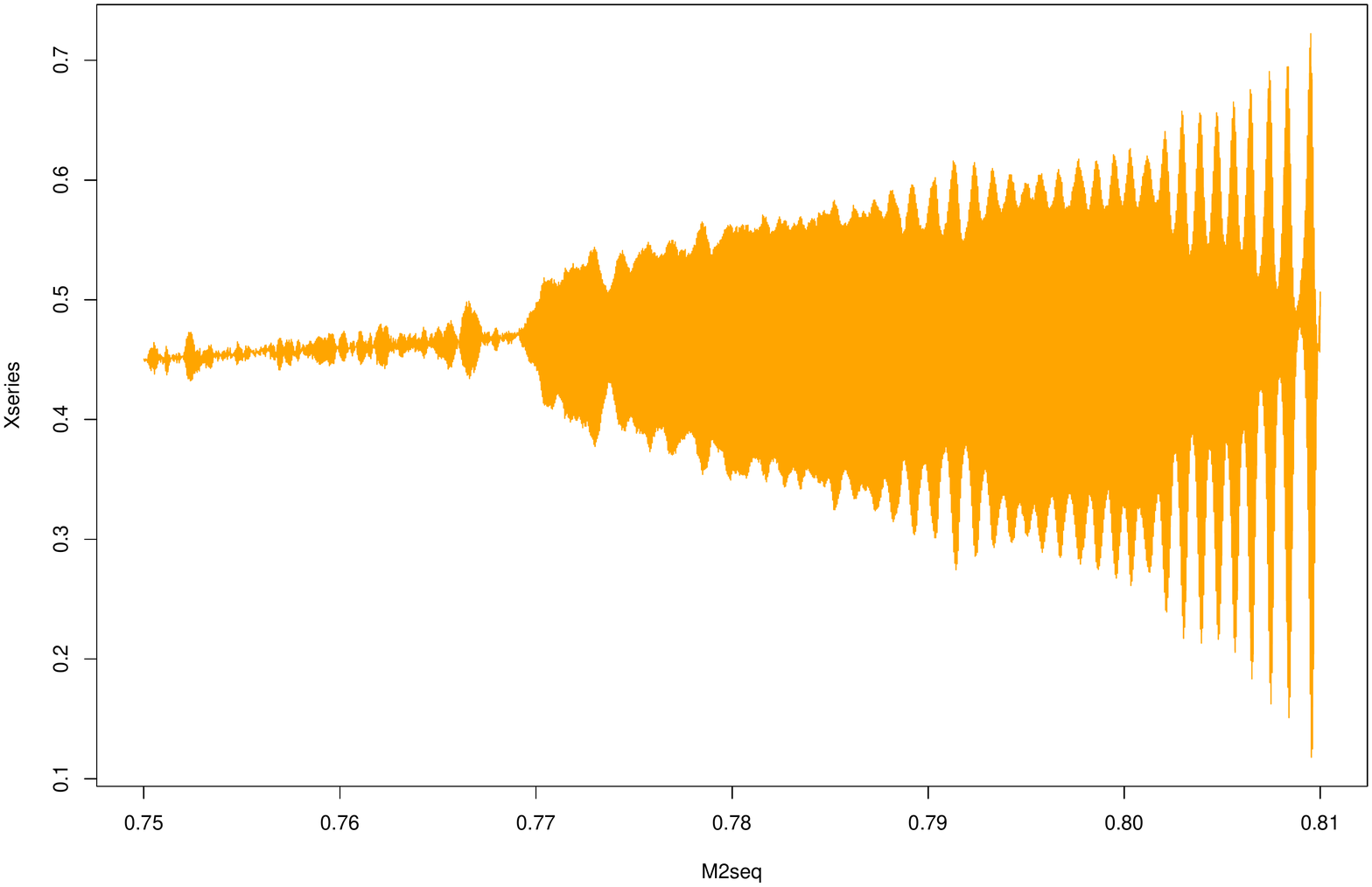}
\caption{$x$-time series for the noisy Lorenz-type system.}
\label{fig:Lorenz-ts}
\end{figure}

We note that, starting with some initial condition $(x_0,y_0,z_0)$ and some initial parameter value of $M_2$, the $x$-time series generated by \eqref{eqn:Lorenz_slow}, shown in Fig.~\ref{fig:Lorenz-ts}, follows closely, but not exactly, the bifurcation diagram in Fig.~\ref{fig:Lorenz_1}, generated by \eqref{eqn:Lorenz}.
The reason for the difference is that at each step of the iteration, the attractor for the corresponding $M_{2,t}$ changes a little, and the one-step iteration of the point $(x_t,y_t,z_t)$ keeps lagging behind the attractor, whose parameter has now changed to $M_{2,t+1}$.

\begin{figure}
\centering
$\begin{array}{cc}
\includegraphics[width=0.5\textwidth]{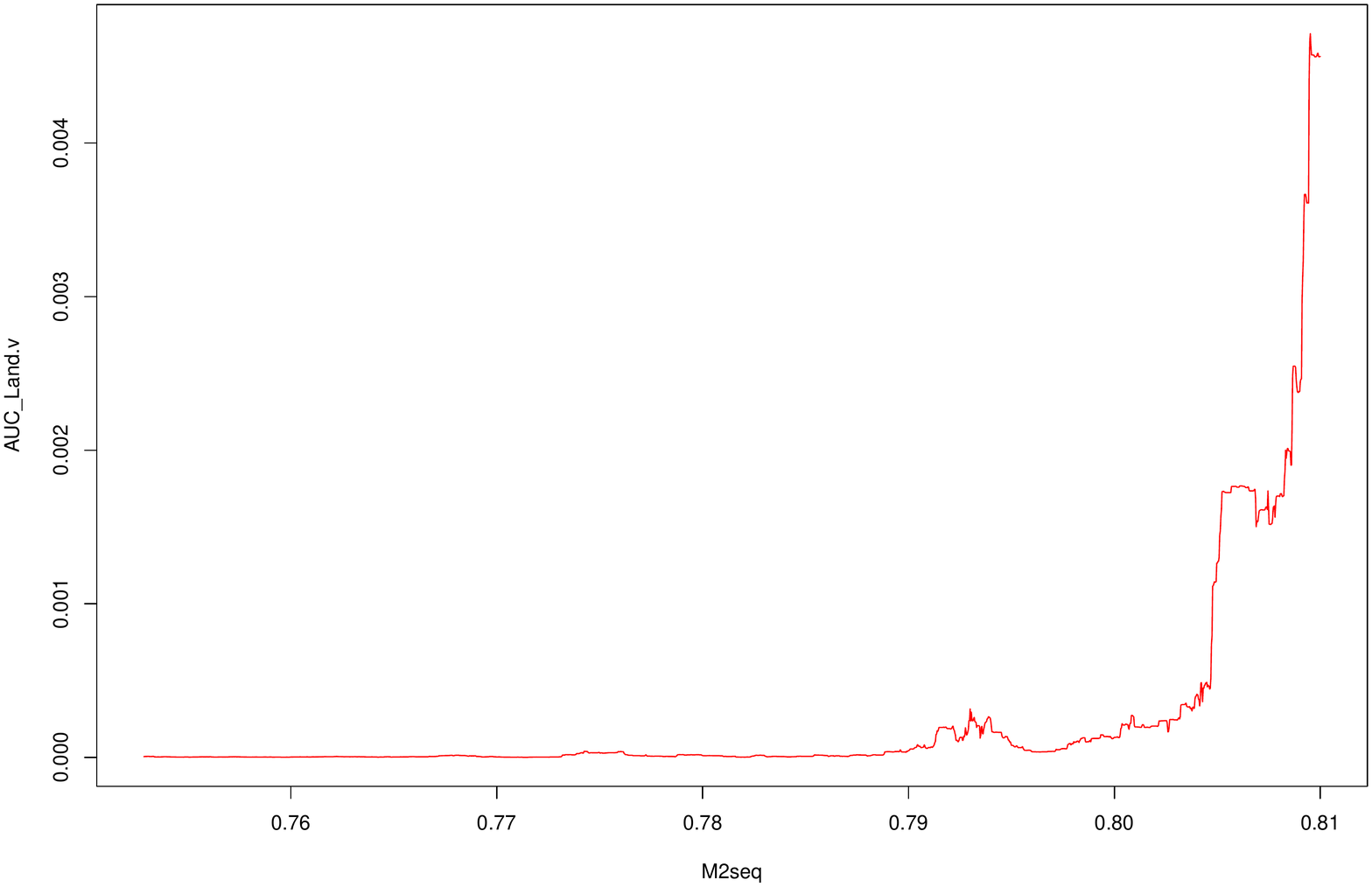}&
\includegraphics[width=0.5\textwidth]{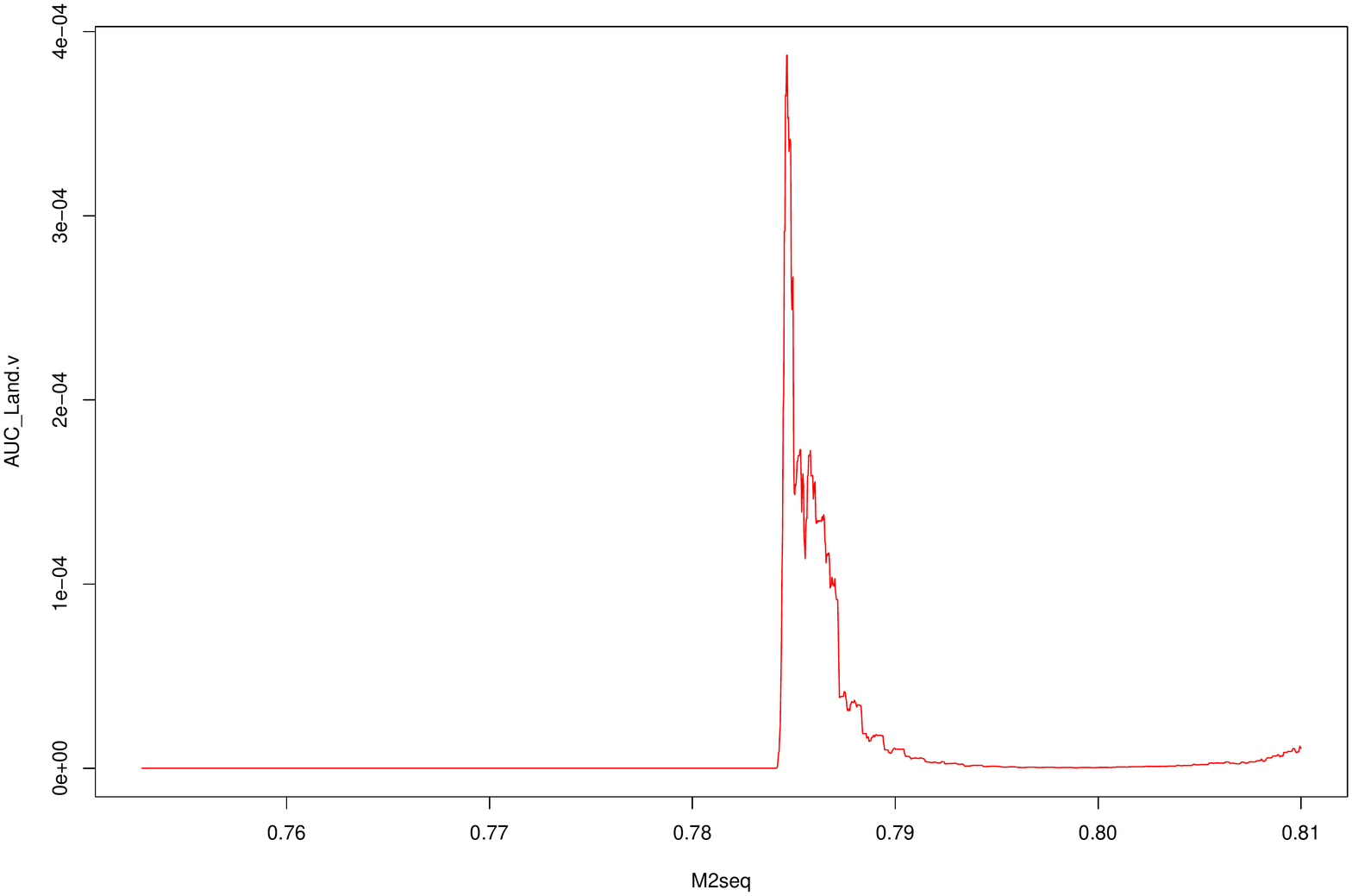}
\end{array}$
\caption{Left: (a) $L^1$-norms of the persistence diagrams for the noisy Lorenz-type system;
Right: (b) $L^1$-norms of the persistence diagrams for the  Lorenz-type system without noise.}
\label{fig:Lorenz-L1}
\end{figure}

As we are interested in critical transition, we cut the time series at  $M_2= 0.81$, which  is approximately the parameter value corresponding to the  bifurcation to a chaotic attractor with frozen parameters; see Fig.~\ref{fig:Lorenz_1}.
As before, we extract the $x$-time series, and we
we use $d=4$ dimensional delay coordinate vectors to reconstruct the phase space.
The total length of the time series is $N=2100$ points.
We use a sliding window of width $w=100$ on the sequence of   $4$-dimensional vectors to generate a sequence of $4$-dimensional point clouds, each point cloud consisting of $100$ points.  This yields $2000$ point-cloud data sets.
On this sequence of cloud points we apply
TDA,  obtaining  the time series of the $L^1$-norms of the $1$-dimensional persistence  landscapes. See Fig.~\ref{fig:Lorenz-L1} (a).

As discussed in Section \ref{sec:reconstruction}, the parameter increment $\Delta M_2$
has to be chosen small enough so that the successive `instantaneous attractors' are very close to one another.
In the above experiment, the smallness condition is that, for the system \eqref{eqn:Lorenz_slow} without noise,  the difference between the $L^1$-norms of successive persistence  landscapes is   less than $\delta=10^{-4}$ in absolute value.

For comparison, we also perform the same computation for the Lorenz-type system without noise, for the same parameter range. See Fig.~\ref{fig:Lorenz-L1} (b). It is striking that for the system without noise there is basically no increase of the $L^1$-norms  before the critical transition, while for the noisy system there is a sharp increase of the $L^1$-norms prior to $M_2=0.81$, although the added noise is small $\epsilon=10^{-3}$.

\begin{figure}
\centering
\includegraphics[width=0.6\textwidth]{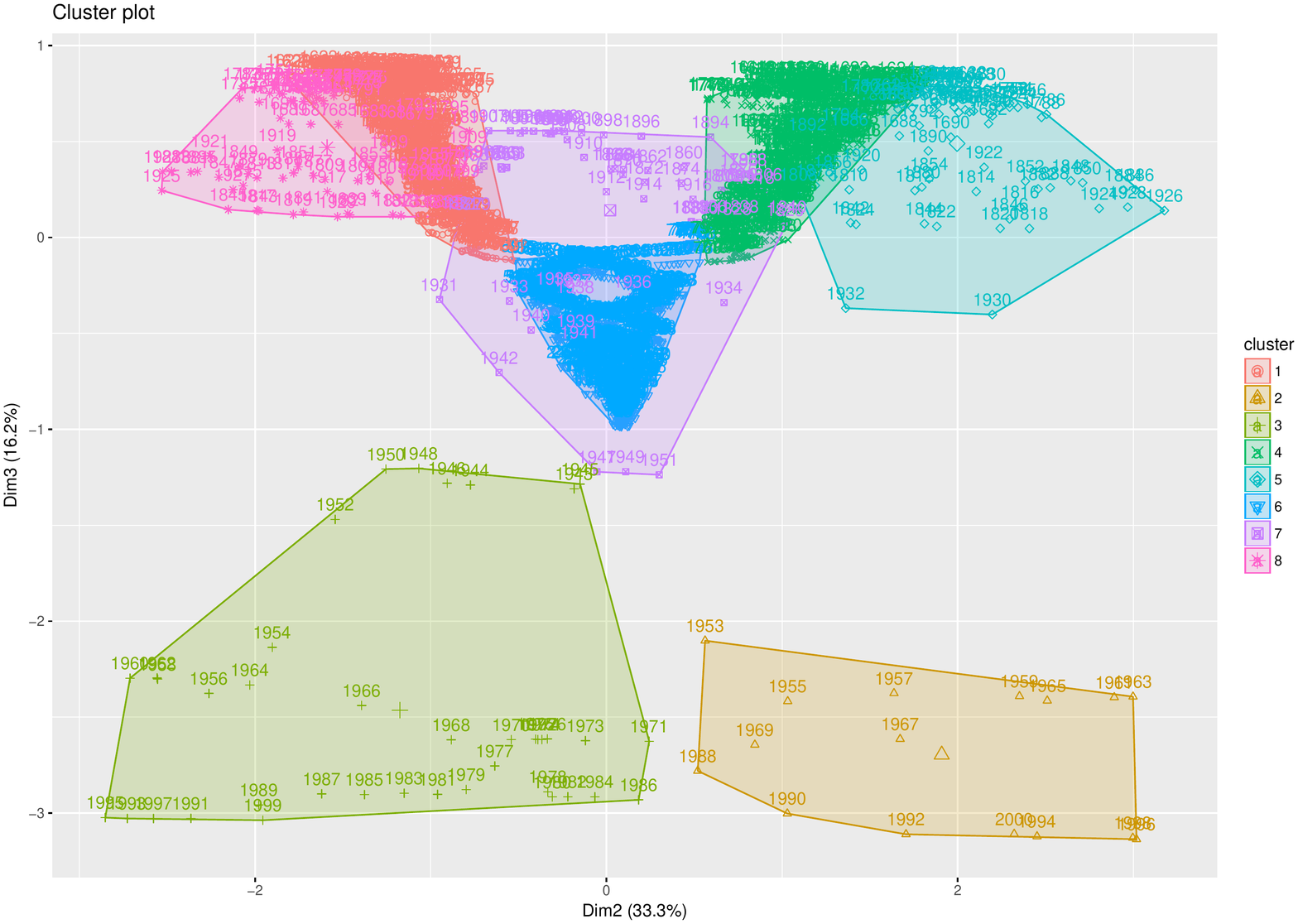}
\caption{$k$-means clustering for the noisy Lorenz-type system time series.}
\label{fig:Lorenz-k}
\end{figure}

To capture the relation between the $x$-time series and the  $L^1$-norms, we use the $k$-means clustering classifier   described in Section \ref{section:k-means}. As input  we use the values of the $x$-time series, and the $L^1$-norms of the persistence landscapes. In order to have a finer segmentation of the data, we  use a relatively large number of clusters $k=8$. The clusters are shown in  Fig.~\ref{fig:Lorenz-k}: clusters 2 and 3 appear most significant, as they are clearly separated from the others. The corresponding data is given in  Table~\ref{eqn:lorenz_c}.
The parameter $M_2$ corresponding to these clusters  ranges between $0.80837$ and $0.80997$.

The conclusion  of this experiments is that  TDA is able to detect the critical transition even though the size of the sliding window (hence the size of the underlying point cloud) is quite small.  The time series of $L^1$-norms grows with the increasing `turbulence' in the noisy system  even prior to the bifurcation value. This is a well known type of behavior in the theory of critical transitions,  where the instability of the system is amplified by the addition of small noise, thus providing early signs of critical transition (see, e.g., \cite{Scheffer2012}).

\section{Analysis of cryptocurrencies}\label{sec:cryptocurrencies}
Cryptocurrencies constitute a new type of financial assets. The underlying  block-chain technology is robust.
Nevertheless, the time series of the relevant exchange rates  feature wild swings in value.  Statistical properties of such assets show strong deviation of distributions of log-returns from normality, and substantial variability of first and second moments in time, i.e. distinctly non-stationary results. These behaviors offer  an interesting real-world data set for studying  `turbulent' behavior, and an excellent test case for applying  TDA for possible detection of critical transitions in financial markets.

We analyze 4 cryptocurrencies -- Bitcoin, Ethereum, Litecoin, and Ripple -- over a period of time from the beginning of 2016 to early 2018. Each of these cryptocurrencies suffered one or two `crashes', during December 2017 -- January 2018.   We apply TDA to the time series of the log of prices (in USD) for each of the 4 cryptocurrencies, and investigate whether the TDA method is able to recognize changes in the relevant time series prior to these crashes. Therefore, for each of the 4 cryptocurrencies under consideration we identify the one or two most significant peaks in the value prior to the crash, and we cut the time series right at the top of the last peak.
The cut-off dates for the 4 time series are slightly different. Also, while Bitcoin and Ethereum show two significant peaks before the crash,  Litecoin and Ripple show only one peak.
To be precise:
\[\begin{tabular}{|l|l|l|l|}
  \hline
  Asset & Starting date & 1st peak & 2nd peak  \\
\hline
  Bitcoin & 2016-01-01 &  2017-12-17            &  2018-01-07  \\
  Ethereum & 2016-01-01 & 2018-01-13           & 2018-01-28    \\
  Litecoin & 2016-01-01  & 2017-12-18  & \\
  Ripple & 2016-01-01  & 2018-01-03  &  \\
  \hline
\end{tabular}
\]

The input for the TDA analysis is the time series of the log-returns for each asset. We perform time-delay coordinate embedding of the time series  as in Section \ref{sec:reconstruction}, with the dimension set to  $d=4$,   and  we apply a sliding window of size  $w=50$. The output of the TDA   is the $L^1$-norm of the persistence landscapes.

\begin{figure}
\centering
\includegraphics[width=0.65\textwidth]{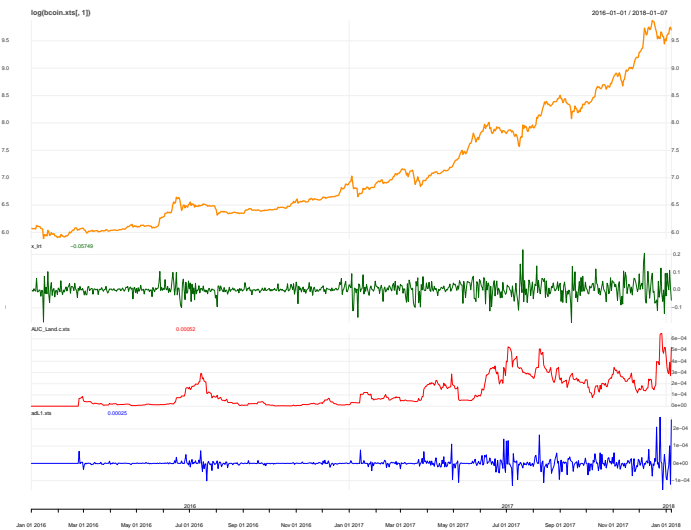}
\caption{Bitcoin 2016-01-01   --  2018-01-07: $\log$ of the values of the asset, the $\log$-return of the asset, the $L^1$-norm, and the first difference of the $L^1$-norms.}
\label{fig:bitcoin-ts}
\end{figure}

\begin{figure}
\centering
\includegraphics[width=0.65\textwidth]{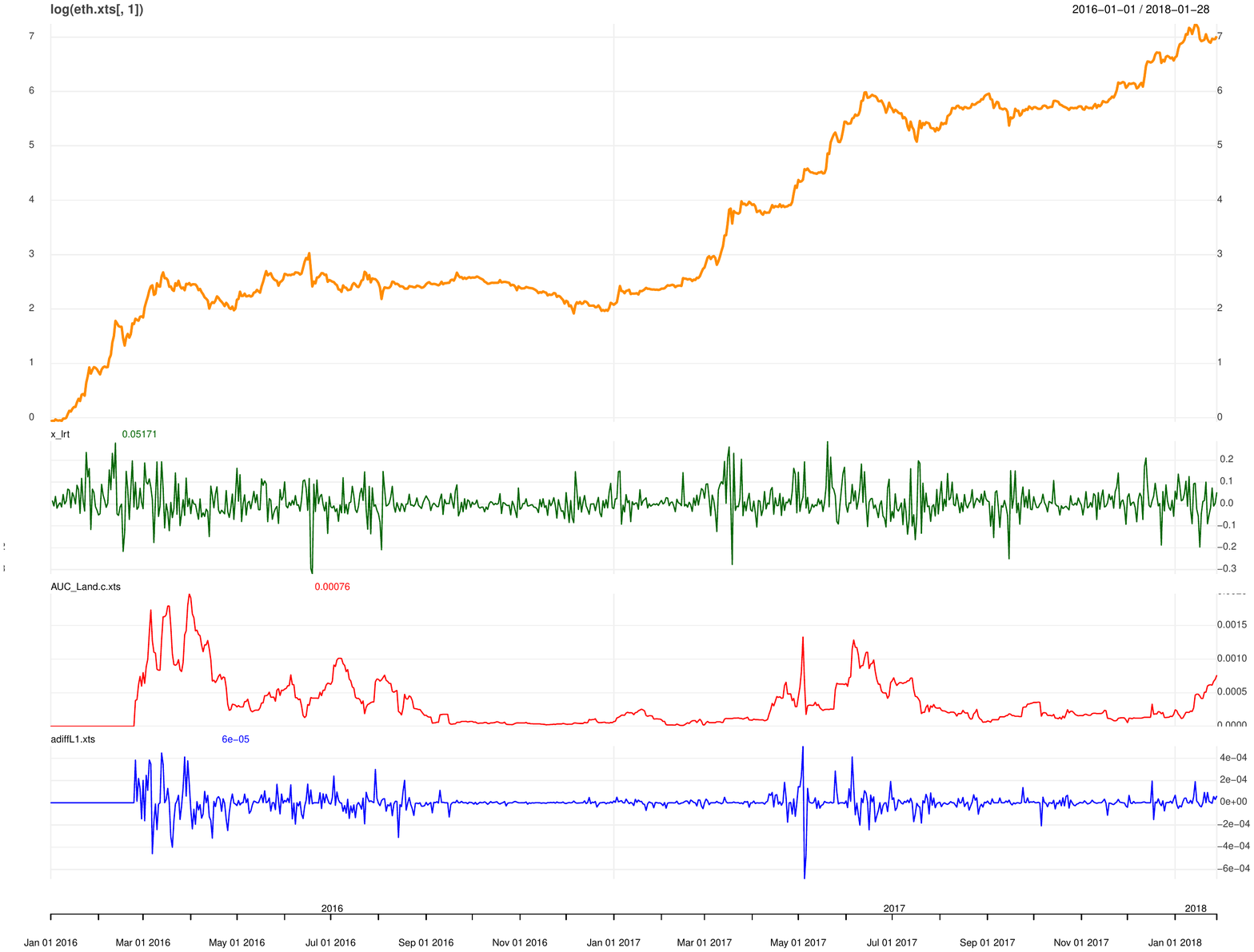}
\caption{Ethereum 2016-01-01 -- 2018-01-28: $\log$ of the values of the asset, the $\log$-return of the asset, the $L^1$-norm, and the first difference of the $L^1$-norms.}
\label{fig:ethereum-ts}
\end{figure}

\begin{figure}
\centering
\includegraphics[width=0.65\textwidth]{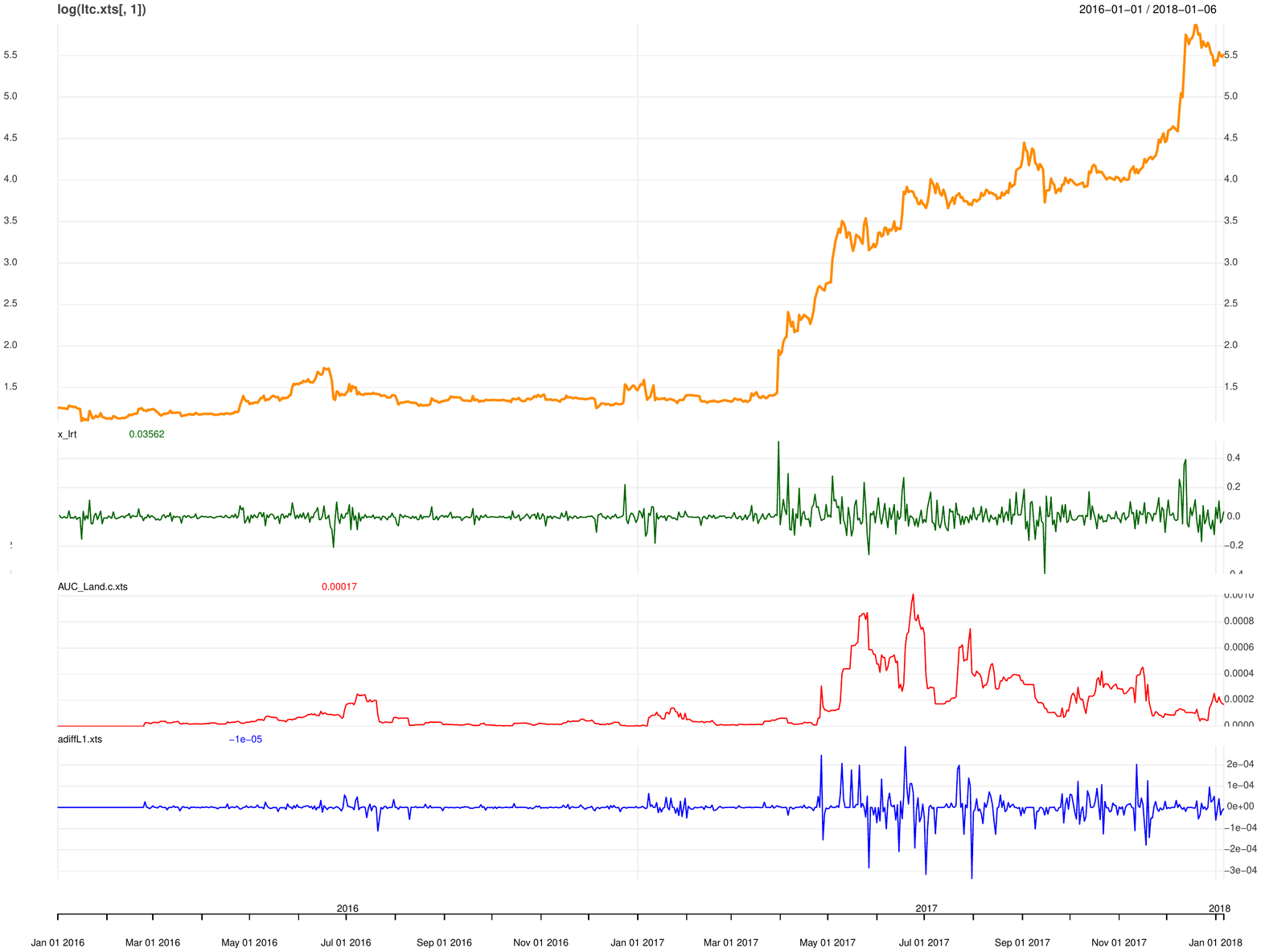}
\caption{Litecoin 2016-01-01  -- 2018-01-06: $\log$ of the values of the asset, the $\log$-return of the asset, the $L^1$-norm, and the first difference of the $L^1$-norms.}
\label{fig:litecoin-ts}
\end{figure}

\begin{figure}
\centering
\includegraphics[width=0.65\textwidth]{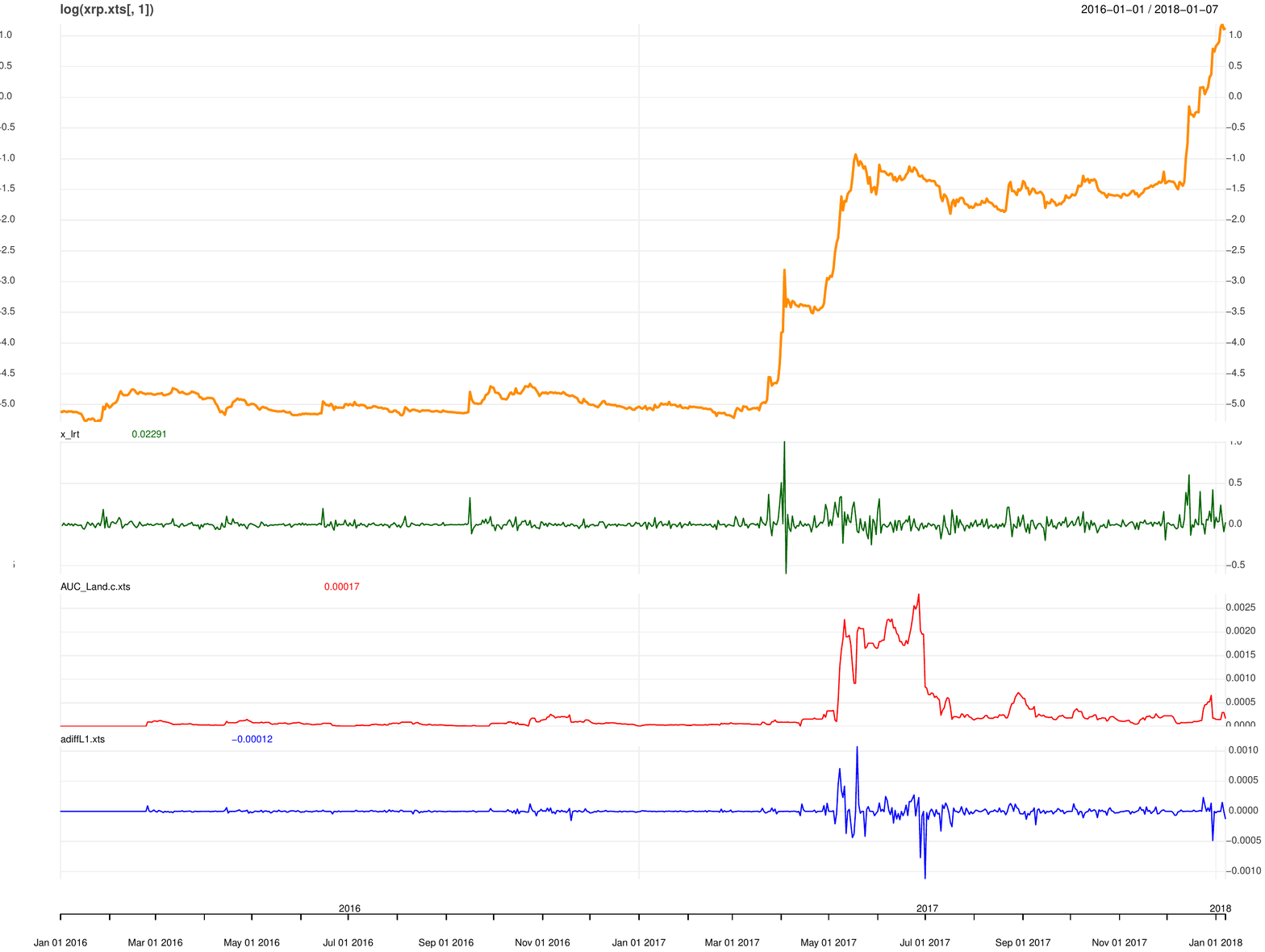}
\caption{Ripple 2016-01-01  -- 2018-01-07: $\log$ of the values of the asset, the $\log$-return of the asset, the $L^1$-norm, and the first difference of the $L^1$-norms.}
\label{fig:ripple-ts}
\end{figure}

We apply TDA and compute the $L^1$-norm of the persistence landscapes.  We display together the $\log$ of the price of the asset, the $\log$-return of the asset, the $L^1$-norm of the persistence landscapes, and the first difference of the $L^1$-norms, for each of the 4 assets, in Fig.~\ref{fig:bitcoin-ts}, Fig.~\ref{fig:ethereum-ts}, Fig.~\ref{fig:litecoin-ts}, Fig.~\ref{fig:ripple-ts}. Inspecting these plots,  we notice that $L^1$-norms tend  to peak in the vicinity of the crashes; there is also an increase in the first difference of the $L^1$-norms.  There are other regions in the time series where the  assets exhibit  large swings, which appear to match  peaks in the  $L^1$-norms; some of these peaks in the  $L^1$-norms are even larger than those prior to the major crash. This makes it clear that we cannot rely on the $L^1$-norms alone to recognize approaching major crashes.


To quantify the relationship between the price/return of each asset  and the $L^1$-norms prior to the crash, we use the unsupervised machine learning technique utilizing non-parametric,  geometry based $k$-means clustering, briefly described in Section \ref{section:k-means}. The main benefit is that this method requires no  statistical assumptions to be satisfied beforehand, in contrast to most statistical tests.

\begin{figure}\label{fig:k}
\centering
$\begin{array}{cc}
\includegraphics[width=0.5\textwidth]{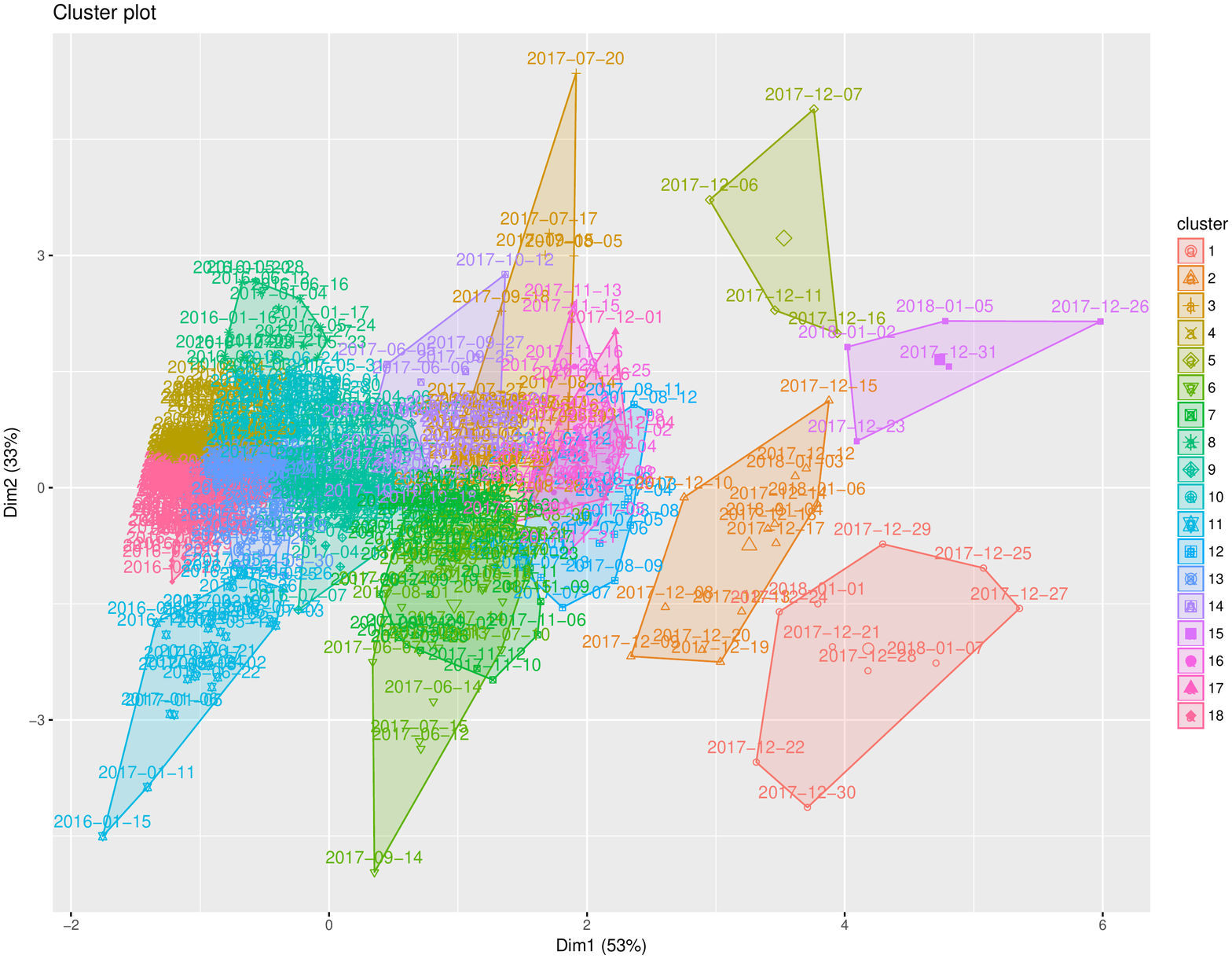}
&\includegraphics[width=0.5\textwidth]{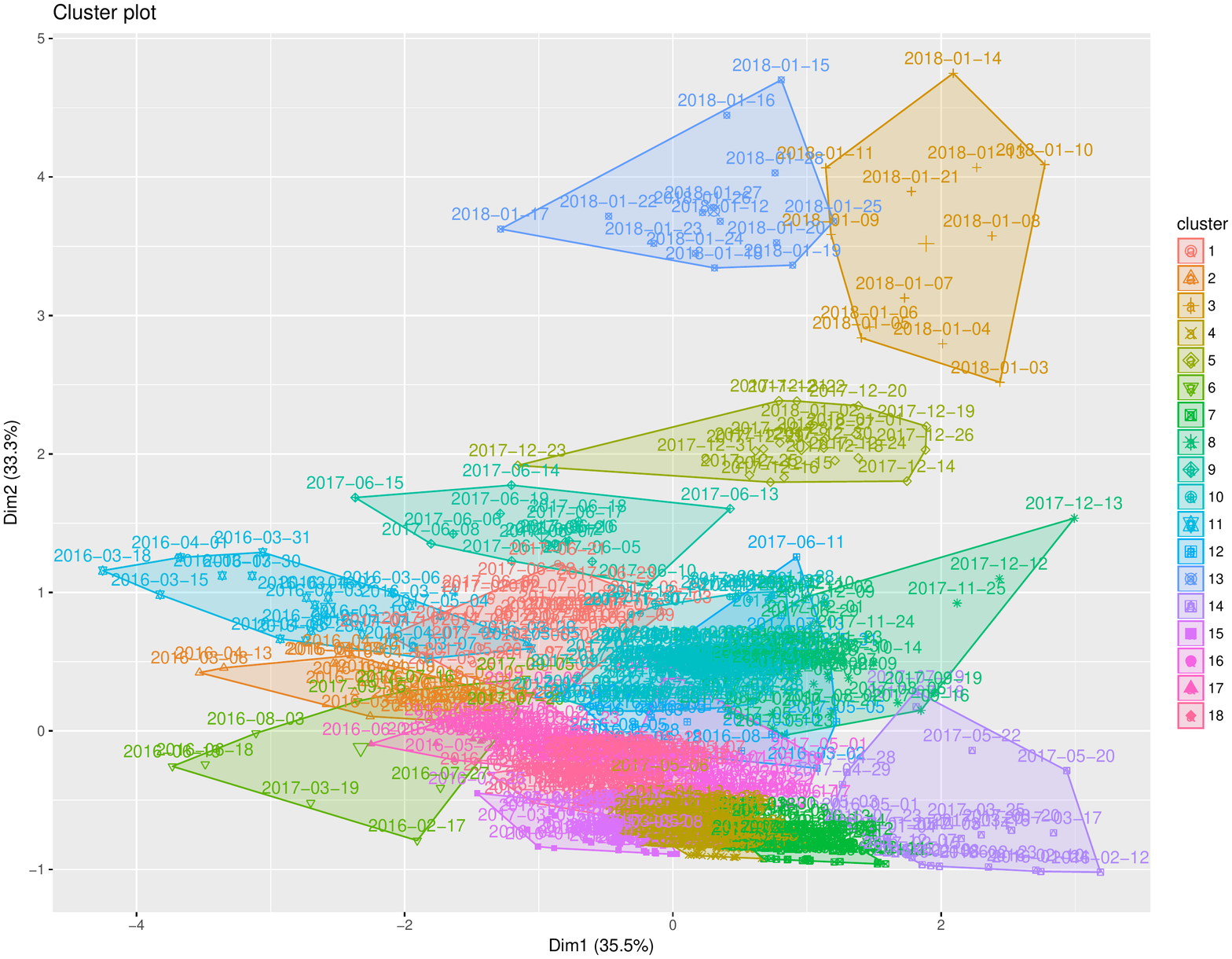}\\
\includegraphics[width=0.5\textwidth]{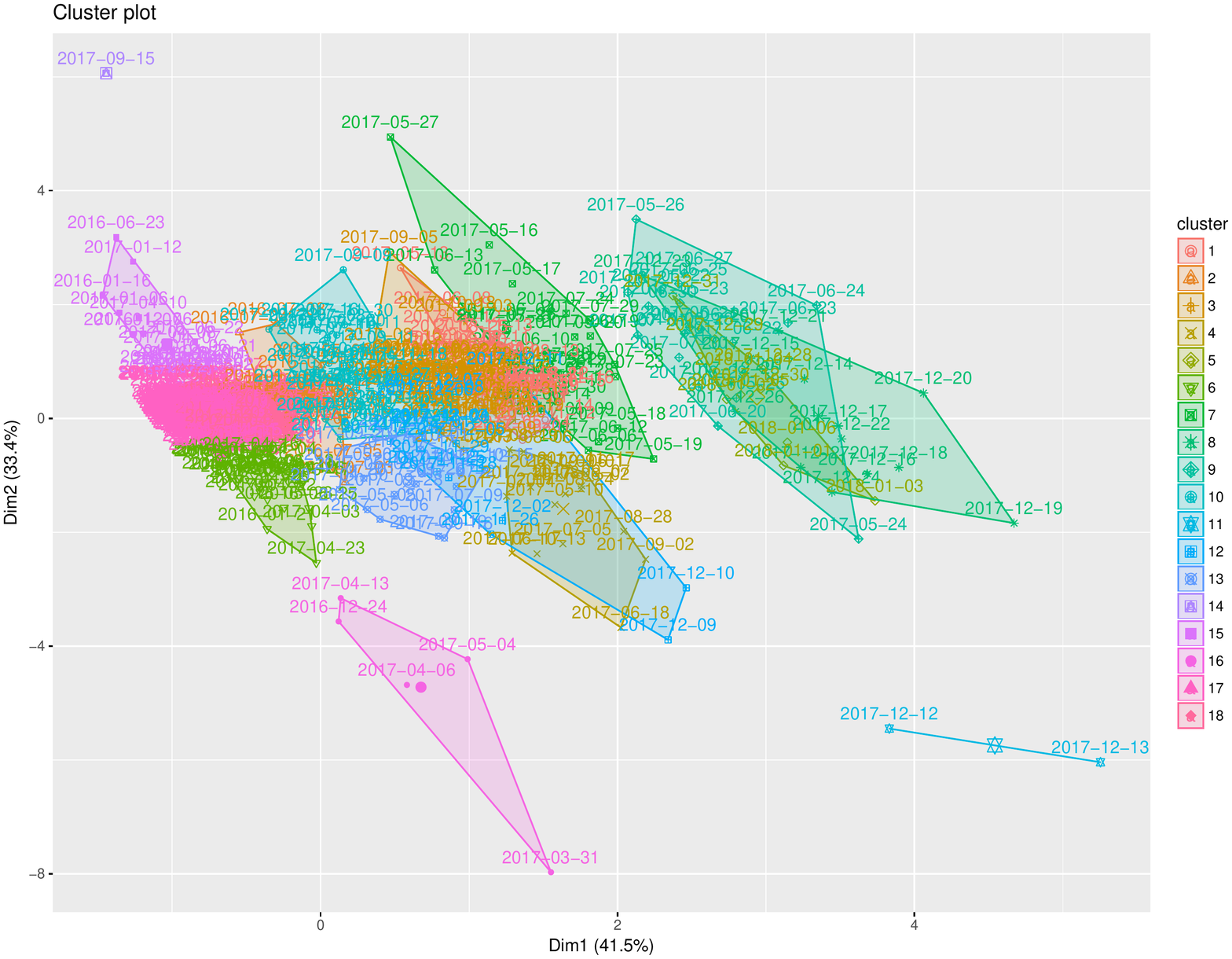}&
\includegraphics[width=0.5\textwidth]{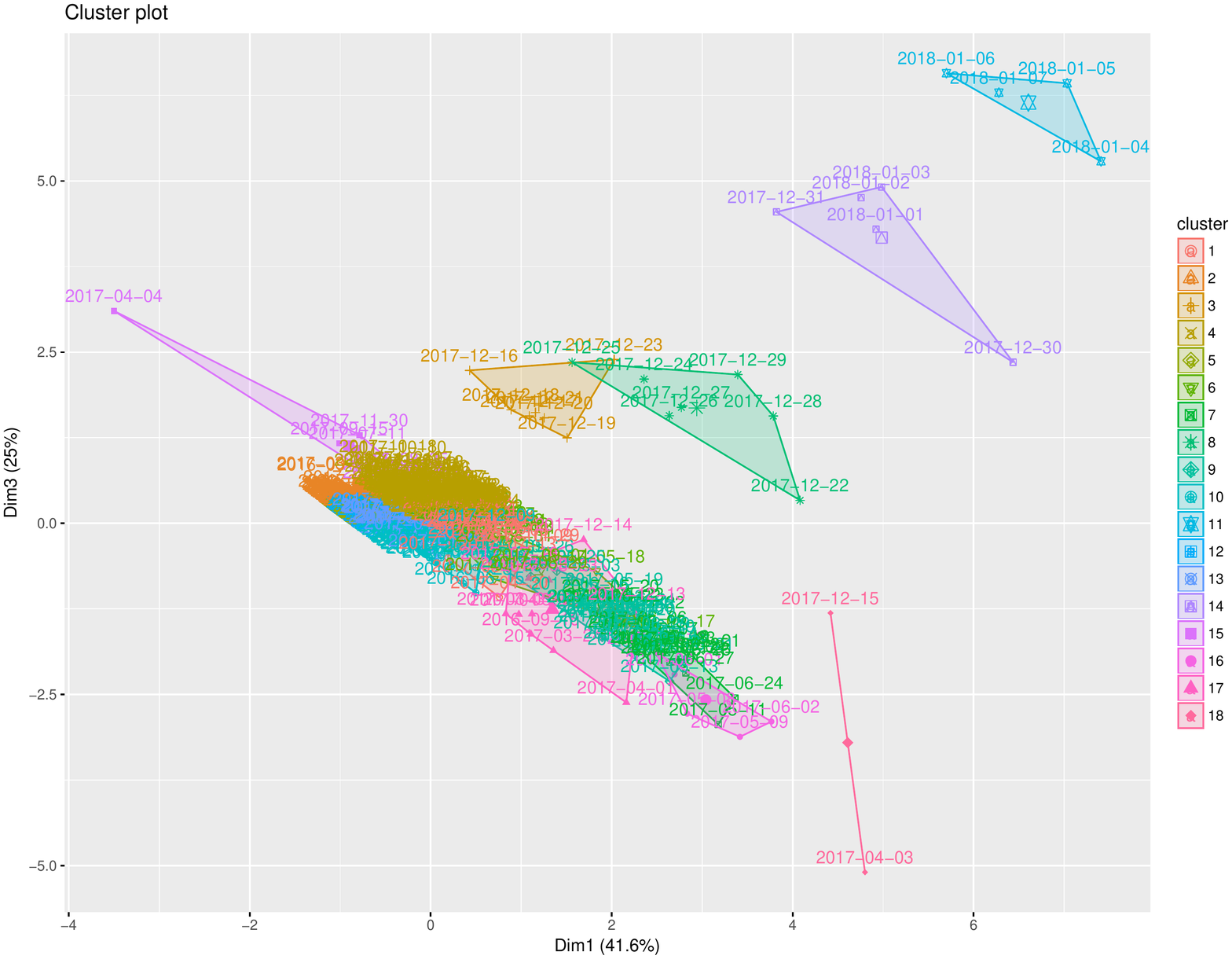}
\end{array}$
\caption{Upper-left panel:  Bitcoin $k$-means clustering:  projections onto principal components  $(1,2)$.
Upper-right panel: Ethereum $k$-means clustering:  projections onto principal components  $(1,2)$.
Lower-left panel: Litecoin $k$-means clustering:  projections onto principal components $(1,2)$.
Lower-right panel: Ripple $k$-means clustering:  projections onto principal components $(1,3)$.}
\end{figure}

As input data we use $x=$ $\log$ of the price of the asset (in USD), $y=$  $\log$-return of the asset, and $z=$  $L^1$-norm of the persistence landscape. Thus, our data set consists of points $(x,y,z)\in\mathbb{R}^3$. Each of these times series is normalized to the $[0,1]$ range.  We  make an empirical choice of  a relatively large number of clusters ($k=18$), which is bigger than the optimal value obtained from the basic methods: `elbow', `silhouette' and `gap statistic'.  The reason of choosing a large $k$ is to obtain a finer segmentation of the data. Smaller values of $k$ yield periods of time that are too large (e.g., larger than the window size  $w=50$), while larger $k$'s
yield periods of time that are too small (e.g., a couple of days).
The clusters are visually depicted by  first performing principal component analysis (PCA) and then projecting onto the principal axes. Thus, for each data set we have 3 possible projections.  In  Fig.~\ref{fig:k} we only show one of the projections which helps us to better identify a small number of significant clusters that are most separated from the others. We stress that PCA plays no role in the $k$-means clustering, as it is only used for the graphical rendering.

The corresponding information on these clusters is given in Table~\ref{eqn:btc_c},  Table~\ref{eqn:eth_c}, Table~\ref{eqn:ltc_c}, Table~\ref{eqn:xrp_c}. (Note that the number associated to each cluster is just an identifier assigned automatically.)
We note  here that these clusters are relatively robust, since running the $k$-means clustering algorithm for slightly lower or bigger values of $k$ leaves these clusters unchanged.

Now we analyze the clusters for the 4 assets:

For Bitcoin, in Table~\ref{eqn:btc_c}, we distinguish two significant clusters. Cluster no. 2 consists of a  discontiguous period between December  8, 2017 -- January 6, 2018. Within this period, the $L^1$-norms display an increasing trend  between December  8, 2017 -- December 15, 2017 -- which is just before the 1-st peak of the Bitcoin on December 17, 2017 --, and a second increasing trend between December 18, 2017 -- January 3, 2018 -- which is just before the 2-nd peak of the Bitcoin on  January 7, 2018. Cluster no. 1 consists of a discontiguous period between December  21, 2017 -- January 7, 2018. Within this period, the $L^1$-norms display an overall increasing trend, up to
the 2-nd peak of the Bitcoin. Overall, the combined information from these two clusters  correctly  identifies the critical period  before the Bitcoin crash, and within that period the $L^1$-norm shows increasing trends that reach their maximum both prior to the 1-st peak  and prior to the 2-nd peak.

For Ethereum, in Table~\ref{eqn:eth_c}, we also distinguish two significant clusters. Cluster no. 3 consists of a discontiguous period between January  3, 2018 -- January 21, 2018. Within this period, the $L^1$-norms displays an overall increasing trend, with a sharp increase around the 1-st peak of the Ethereum on January 13, 2018.
Cluster no. 13 consists of a  discontiguous period between January  12, 2018 -- January 28, 2018, before the 2-nd  peak of the Ethereum. The $L^1$-norms follow an increasing trend throughout the whole period.

For Litecoin, in Table~\ref{eqn:ltc_c}, we  distinguish one significant cluster. Cluster no. 12 consists of a discontiguous period between November 23, 2017 -- December 11, 2018. During this period the $L^1$-norms show an increasing trend.   We note here that this increasing trend continues  during the period  December 12, 2017 -- December 13, 2017 -- which is part of cluster no 11 (not shown in the table) --,
and  plateaus during  December 14, 2017 -- December 18, 2017,  -- which is part of cluster 8 (not shown in the table). Thus, the increasing trend occurs over a significant period prior the  peak of Litecoin on December 18, 2017, even though this happens across multiple clusters.

For Ripple, in Table~\ref{eqn:xrp_c},  we  also distinguish one significant cluster. Cluster no 8 consists of one  discontiguous period between December 22, 2017 -- December 29, 2017, where  the $L^1$-norms display a steep increasing. This period precedes the first peak of the Ripple on January 3, 2018.

Thus, the $k$-means clustering applied to  the data consisting of the log-price of each asset, the log-return, and the $L^1$-norms of the persistence landscapes,  provides  identification of topologically distinct regimes before the crash of each asset. Within each corresponding period of time, we make an empirical observation that  the $L^1$-norms follow an increasing trend, a behavior consistent with the findings in \cite{GideaKatz18}.


\section{Conclusions}\label{section:conclusion} We have presented a TDA-based method to detect critical transitions
in complex systems by exploring changes in the topological properties of the relevant
time series. The proposed method involves time-delay coordinate embedding, sliding windows, and persistence landscapes. It can be applied to systems that are strongly nonlinear, and performs well even with short-time windows, whereas   typical statistical methods are prone to ambiguities. We applied this method to the time series of the most  capitalized
cryptocurrencies -- Bitcoin, Ethereum, Litecoin, and Ripple  -- for some interval of time before the  crash at the end of 2017 -- beginning 2018. Furthermore, we have used a $k$-means clustering technique to identify topologically distinct regimes in the time series of observed and derived signals for each cryptocurrency, and matched them  to relevant time periods prior to the crash.  In summary, we conclude that the combination of the TDA-based method and $k$-means clustering technique has the potential to automatically recognize approaching critical transitions in the cryptocurrency markets, even when the relevant time series  exhibit  a highly non-stationary, erratic behavior.

\section{Acknowledgement}
Research of M.G. was partially supported by NSF grant  DMS-0635607,  NSF grant  DMS-1814543, and by the  Alfred P. Sloan Foundation grant G-2016-7320. Research of Y.K. was supported by S\&P Global Market Intelligence. Research of P.R. was partially supported by NSF grant  DMS-1814543.
The views expressed in this paper are those of the authors, and do not necessary represent the views of S\&P Global Market Intelligence.

\section*{Appendix}

\begin{tab}\label{eqn:lorenz_c}{Lorenz-type attractor -- significant cluster(s)}
\begin{equation*}
\begin{split}
&\begin{array}[t]{cc}
\begin{filecontents*}{Lorenz_k8_c2b.csv}
Cluster 2,x,L1
0.80866,0.42953,0.77040
0.80871,0.34612,0.83859
0.80877,0.24360,0.82747
0.80883,0.12184,0.82854
0.80889,0.02995,0.82820
0.80894,0.01178,0.82759
0.80900,0.09461,0.83469
0.80906,0.23495,0.88262
0.80911,0.37531,0.89304
0.80966,0.42885,0.92943
0.80971,0.33974,0.97690
0.80977,0.22359,0.99851
0.80983,0.09652,0.99851
0.80989,0.00000,1.00000
0.80994,0.00361,0.99901
0.81000,0.11870,0.99780
 ,,
 ,,
 ,,
 ,,
 ,,
 ,,
 ,,
 ,,
 ,,
 ,,
 ,,
 ,,
 ,,
 ,,
 ,,
 ,,
 ,,
 ,,
 ,,
 ,,
 ,,
 ,,
 ,,
\end{filecontents*}
\csvautotabular{Lorenz_k8_c2b.csv} &\begin{filecontents*}{Lorenz_k8_c3.csv}
Cluster 3,x,L1
0.80837,0.56476,0.59651
0.80840,0.66550,0.59474
0.80843,0.55931,0.59136
0.80846,0.68811,0.59383
0.80851,0.71626,0.57794
0.80857,0.74806,0.58001
0.80863,0.79444,0.63943
0.80869,0.84840,0.78818
0.80874,0.90712,0.84313
0.80880,0.95758,0.82854
0.80886,0.98429,0.82854
0.80891,0.95804,0.82818
0.80897,0.86784,0.83469
0.80903,0.75833,0.85612
0.80909,0.66972,0.89405
0.80914,0.61150,0.89304
0.80917,0.47779,0.89203
0.80920,0.58578,0.89292
0.80923,0.53950,0.89331
0.80926,0.58187,0.89325
0.80929,0.58795,0.89325
0.80931,0.57645,0.89296
0.80934,0.62589,0.92549
0.80937,0.57304,0.95414
0.80940,0.65224,0.95368
0.80943,0.56844,0.96050
0.80946,0.67968,0.96050
0.80949,0.55336,0.96050
0.80951,0.71205,0.96050
0.80954,0.52710,0.96050
0.80957,0.75040,0.96352
0.80960,0.48456,0.96352
0.80963,0.79184,0.96412
0.80969,0.85220,0.97871
0.80974,0.91706,0.99640
0.80980,0.97867,0.99851
0.80986,1.00000,0.99851
0.80991,0.95320,0.99901
0.80997,0.84711,0.99901
\end{filecontents*}
\csvautotabular{Lorenz_k8_c3.csv}
\end{array}\end{split}
\end{equation*}
\end{tab}

\begin{tab}\label{eqn:btc_c}{Bitcoin -- significant cluster(s)}
\begin{equation*}
\begin{split}
&\begin{array}[t]{cc}
\begin{filecontents*}{btc_c2.csv}
Cluster 2,price,return,L1
12/8/2017,0.827,0.328,0.209
12/9/2017,0.767,0.266,0.225
12/10/2017,0.773,0.467,0.247
12/12/2017,0.886,0.516,0.373
12/13/2017,0.845,0.335,0.354
12/14/2017,0.852,0.465,0.371
12/15/2017,0.908,0.6,0.362
12/17/2017,0.986,0.414,0.244
12/18/2017,0.98,0.431,0.226
12/19/2017,0.909,0.266,0.27
12/20/2017,0.848,0.281,0.288
1/3/2018,0.778,0.511,0.467
1/4/2018,0.779,0.45,0.447
1/6/2018,0.884,0.475,0.418
\end{filecontents*}
\csvautotabular{btc_c2.csv} &\begin{filecontents*}{btc_c1.csv}
Cluster 1,price,return,L1
12/21/2017,0.801,0.311,0.613
12/22/2017,0.711,0.164,0.606
12/24/2017,0.717,0.349,0.568
12/25/2017,0.714,0.437,0.983
12/27/2017,0.791,0.389,1
12/28/2017,0.741,0.291,0.763
12/29/2017,0.741,0.447,0.727
12/30/2017,0.646,0.122,0.806
1/1/2018,0.688,0.367,0.677
1/7/2018,0.833,0.307,0.807
,,,
,,,
,,,
  ,,,
\end{filecontents*}
\csvautotabular{btc_c1.csv}
\end{array}\end{split}
\end{equation*}
\end{tab}
\begin{tab}\label{eqn:eth_c}{Ethereum: significant cluster(s)}
\begin{equation*}\begin{split}
&\begin{array}[t]{cc}
\begin{filecontents*}{eth_c3.csv}
Cluster 3 ,price,return,L1
1/3/2018,0.634,0.754,0.093
1/4/2018,0.688,0.663,0.079
1/5/2018,0.698,0.552,0.06
1/6/2018,0.712,0.56,0.064
1/7/2018,0.746,0.605,0.085
1/8/2018,0.829,0.701,0.106
1/9/2018,0.82,0.51,0.107
1/10/2018,0.93,0.736,0.105
1/11/2018,0.907,0.486,0.116
1/13/2018,0.909,0.68,0.139
1/14/2018,1,0.685,0.236
1/21/2018,0.827,0.694,0.258
,,,
 ,,,
\end{filecontents*}
\csvautotabular{eth_c3.csv} &\begin{filecontents*}{eth_c13.csv}
Cluster 13,price,return,L1
1/12/2018,0.829,0.378,0.112
1/15/2018,0.977,0.489,0.243
1/16/2018,0.925,0.438,0.243
1/17/2018,0.759,0.202,0.229
1/18/2018,0.727,0.456,0.21
1/19/2018,0.736,0.548,0.21
1/20/2018,0.748,0.553,0.258
1/22/2018,0.755,0.378,0.306
1/23/2018,0.718,0.446,0.312
1/24/2018,0.706,0.5,0.314
1/25/2018,0.761,0.65,0.311
1/26/2018,0.753,0.511,0.34
1/27/2018,0.755,0.532,0.355
1/28/2018,0.795,0.613,0.384
\end{filecontents*}
\csvautotabular{eth_c13.csv}
\end{array}\end{split}\end{equation*}\end{tab}
\begin{tab}\label{eqn:eth_c}{Litecoin: significant cluster(s)}
\begin{equation*}\label{eqn:ltc_c}\begin{split}
&\begin{array}[t]{c}
\begin{filecontents*}{ltc_c12.csv}
Cluster 12,price,return,L1
11/23/2017,0.19,0.46,0.08
11/25/2017,0.21,0.5,0.11
11/26/2017,0.24,0.58,0.09
11/28/2017,0.25,0.5,0.07
11/29/2017,0.26,0.48,0.07
12/1/2017,0.24,0.45,0.09
12/2/2017,0.27,0.56,0.09
12/3/2017,0.27,0.44,0.09
12/4/2017,0.28,0.44,0.09
12/5/2017,0.28,0.46,0.09
12/6/2017,0.28,0.41,0.09
12/7/2017,0.27,0.4,0.12
12/8/2017,0.27,0.41,0.12
12/9/2017,0.35,0.72,0.13
12/10/2017,0.43,0.65,0.13
12/11/2017,0.41,0.38,0.12
\end{filecontents*}
\csvautotabular{ltc_c12.csv}
\end{array}\end{split}\end{equation*}\end{tab}
\begin{tab}\label{eqn:xrp_c}{Ripple: significant cluster(s)}
\begin{equation*}\begin{split}
&\begin{array}[t]{c}
\begin{filecontents*}{xrp_c8.csv}
Cluster 8,price,return,L1
12/22/2017,0.35,0.62,0.04
12/24/2017,0.36,0.38,0.13
12/25/2017,0.32,0.30,0.16
12/26/2017,0.34,0.41,0.16
12/27/2017,0.36,0.41,0.19
12/28/2017,0.42,0.47,0.19
12/29/2017,0.44,0.40,0.23
\end{filecontents*}
\csvautotabular{xrp_c8.csv}
\end{array}
\end{split}\end{equation*}
\end{tab}



\begin{thebibliography}{00}
\bibitem{Adler} Adler, R.J.; Bobrowski, O.;   Borman, M.S.; Subag E.;  Weinberger, S. (2010). Persistent homology
for random fields and complexes. In Borrowing Strength: Theory Powering Applications,
A Festschrift for Lawrence D. Brown, IMS Collections 6, 124--143.
\bibitem{BerwaldGidea14} Berwald, J.; Gidea, M. (2014). Critical Transitions in a Model of a Genetic
Regulatory System. Mathematical Biology and Engineering,
11:723, 2014.
\bibitem{BerwaldGideaVejdemo14} Berwald J.; Gidea, M; Vejdemo-Johansson, M. (2014). Automatic Recognition and Tagging of Topologically Different Regimes in Dynamical Systems. Discontinuity, Nonlinearity and Complexity,  Vol. 3.
\bibitem{Bubenik15} Bubenik,  P.  (2015). Statistical topological data analysis using persistence landscapes, J. Machine Learning Research  Vol. 16,  77.
\bibitem{Bubenik16}  Bubenik  P.; Dlotko P. (2016). A persistence landscapes toolbox for topological statistics, Journal of Symbolic Computation, Vol. 78, 91.
\bibitem{Drozdz} Dro{\.z}d{\.z}, S., G{\c e}barowski, R., Minati, L., O{\'s}wi{\c e}cimka, P., \& W{\c a}torek, M.\ (2018). Bitcoin market route to maturity? Evidence from return fluctuations, temporal correlations and multiscaling effects. eprint arXiv:1804.05916
\bibitem{Carlsson09} Carlsson, G. (2009). Topology and Data, Bull. Amer. Math. Soc., Vol. {46}
  255.
\bibitem{Casdagli2002} Casdagli, M.; Eubank, S.;  Farmer, J.D.; Gibson, J (2002). State space reconstruction in the presence of noise.   Physica D: Nonlinear Phenomena, 51(1--3),  52--98.
\bibitem{Catania2018} Catania, L., Grassi, S. and Ravazzolo, F. (2018). Forecasting Cryptocurrencies Financial Time Series (No. 5/2018).
\bibitem{Catania2017} Catania, L.; Grassi, S. (2017). Modelling Crypto-Currencies Financial Time-Series. Available at SSRN: https://ssrn.com/abstract=3028486 or http://dx.doi.org/10.2139/ssrn.3028486
\bibitem{Chan2017} Chan, S.; Chu, J.; Nadarajah, S.;  Osterrieder, J. (2017). A statistical analysis of cryptocurrencies. Journal of Risk and Financial Management, 10(2), p.12.
\bibitem{Chazal2015} Chazal, F.; Fasy, B.T.; Lecci, F.;  Rinaldo, A.;  Wasserman, L. (2015). Stochastic convergence of persistence landscapes and silhouettes, Journal of Computational Geometry Vol. 6, 140--161.
\bibitem{Chiu2017} Chiu, J.; Koeppl, T. (2017). The economics of cryptocurrencies - bitcoin and beyond (No. 6688). Victoria University of Wellington, School of Economics and Finance.
\bibitem{Chu2017} Chu, J.; Chan, S.; Nadarajah, S.; Osterrieder, J. (2017). GARCH Modelling of Cryptocurrencies. Journal of Risk and Financial Management, 10(4), p.17.
\bibitem{Cohen-Steiner} Cohen-Steiner, D.;  Edelsbrunner, H.; Harer, J. (2007). Stability of persistence diagrams,
Discrete \& Computational Geometry, Vol.37, 103.
\bibitem{Dakos2012} Dakos, V.; Carpenter, S.R.; Brock, W.A.; Ellison, A.M.; Guttal, V. et al. (2012)
 Methods for Detecting Early Warnings of Critical Transitions in Time Series Illustrated Using Simulated Ecological Data. PLOS ONE 7(7): e41010
\bibitem{Edelsbrunner10}  Edelsbrunner, H.; Harer, J. (2010). Computational topology: an introduction (Amer. Math. Soc.
\bibitem{Holzmann2008} Holzmann, H.; Vollmer, S. (2008).  A likelihood ratio test for bimodality in two-component mixtures with application to regional income distribution in the EU. AStA Advances in Statistical Analysis. Vol. 2, 57-69.
\bibitem{GerlachSornette} Gerlach, J.-C.; Demos, G.; Sornette, D. (2018).	Dissection of Bitcoin's Multiscale Bubble History from January 2012 to February 2018. eprint arXiv:1804.06261
\bibitem{Garland} Garland, J.; Bradley, E.;  Meiss, J.D. (2016). Exploring the topology of dynamical reconstructions. Physica D: Nonlinear Phenomena, 334, 49--59.
\bibitem{Gidea17} Gidea, M. (2017)  Topological Data Analysis of Critical Transitions in Financial Networks. In International Conference and School on Network Science (pp. 47-59). Springer, Cham.
\bibitem{GideaKatz18} Gidea, M.;  Katz, Y. (2018). Topological data analysis of financial time series: Landscapes of crashes. Physica A: Statistical Mechanics and its Applications, Vol. 491, 820--834.
\bibitem{Gonchenko}  Gonchenko, S.V.;  Ovsyannikov, I.I.; Sim\'o, C.; Turaev, D. (2005) Three-Dimensional H\'enon-Like Maps and Wild Lorenz-Like Attractors. International Journal of Bifurcation and Chaos, Vol. 15, No. 11, 3493--3508.
\bibitem{Guttal2008} Guttal, V.; Jayaprakash, C. (2008). Changing skewness: an early warning signal of regime shifts in ecosystems. Ecology letters, 11(5), pp.450-460.
\bibitem{Jain} Jain, A.K. (2010). Data clustering: 50 years beyond K-means. Pattern recognition letters, 31(8), 651--666.
\bibitem{Kennel} Kennel, M.B.; Brown, R.; Abarbanel, H.D. (1992). Determining embedding dimension for phase-space reconstruction using a geometrical construction. Physical review A, 45(6), p.3403.
\bibitem{Livnia2007} Livina V.N.; Lenton T.M. (2007). A modified method for detecting incipient bifurcations in a dynamical system. Geophys Res Lett 34.
\bibitem{Lum15}   Lum, P.Y.;  Singh, G.;  Lehman, A.;  Ishkanov, T.;  Vejdemo-Johansson, M.;  Alagappan, M.;  Carlsson, J.;    Carlsson, G. (2013) Extracting insights from the shape of complex data using topology. Scientific Reports, Vol. {3},  1236.
\bibitem{Munch2016} Khasawneh, F.A.; Munch, E. (2016).  Chatter detection in turning using persistent homology, Mechanical Systems and Signal Processing, Vol. {70-71},  527.
\bibitem{Khasawneh2018}	Khasawneh, F.A.; Munch, E. (2018). Topological Data Analysis for True Step Detection in Piecewise Constant Signals. eprint arXiv:1805.06403
\bibitem{Kozlowska2016} Kozlowska, M.; Denys, M.; Wilinski, M.; Link, G.; Gubiec, T.; Werner, T.R.; Kutner, R.;  Struzik, Z.R. (2016). Dynamic bifurcations on financial markets. Chaos, Solitons \& Fractals, 88, 126--142.
\bibitem{Livan} Livan, G; Inoue J.  and  Scalas, E (2012). On the non-stationarity of financial time series: impact on optimal portfolio selection.  Journal of Statistical Mechanics: Theory and Experiment, Volume 2012,  P07025.
\bibitem{Lu} Lu, Lerong (2018). Bitcoin: Speculative Bubble, Financial Risk and Regulatory Response. Butterworths Journal of International Banking and Financial Law (2018). 33. 178-182.
\bibitem{Mischaikow-et-al}  Kramar, M.; Levanger,  R.; Tithof, J.; Suri,  B.; Xu,  M.; Paul,  M.;  Schatz, M.F.;   Mischaikow,  K. (2016). Analysis of Kolmogorov flow and Rayleigh-Benard convection using persistent homology. Physica D: Nonlinear Phenomena, Vol. {334},  82.
\bibitem{Muldoon}  Muldoon, M.R.;  MacKay, R.S.;  Huke, J.P.;  Broomhead, D.S. (1993). Topology from time series.
Physica D: Nonlinear Phenomena, Vol. 65,  1--16.
\bibitem{Munix2012} M\"unnix, M. C., Shimada, T., Schäfer, R., Leyvraz, F., Seligman, T. H., Guhr, T., \& Stanley, H. E. (2012). Identifying States of a Financial Market. Scientific Reports, 2, 644. http://doi.org/10.1038/srep00644
\bibitem{Packard} Packard, N.H.,  Crutchfield, J.P., Farmer, J.D.,   Shaw, R.S. (1980) Geometry from a time series. Phys. Rev. Lett. Vol. 45, 712.
\bibitem{Perea} Perea, J.A. and Harer, J. (2015). Sliding windows and persistence: An application of topological methods to signal analysis. Foundations of Computational Mathematics, 15(3),  799--838.
\bibitem{Perea2015} Perea, J.A.; Deckard, A.; Haase, S.B.;  Harer, J.(2015). SW1PerS: Sliding windows and 1-
persistence scoring; discovering periodicity in gene expression time series data. BMC
Bioinformatics, Vol. 16.
\bibitem{Sauer} Sauer, T.; Yorke, J. A.;  Casdagli, M. (1991). Embedology. J. Statistical Physics 65:579.\bibitem{Scheffer2009} Scheffer, M., et al. (2009). Early-warning signals for critical transitions. Nature, 461.
\bibitem{Scheffer2012} Scheffer, M.; Carpenter, S.R.; Lenton, T.M.; Bascompte, J.; Brock, W.; Dakos, V.; Van de Koppel, J.; Van de Leemput, I.A.; Levin, S.A.; Van Nes, E.H.; Pascual, M. (2012). Anticipating critical transitions. Science, 338(6105), pp.344-348.
\bibitem{Seversky} Seversky, L.M.; Davis, S.; Berger, M. (2016). On time-series topological data analysis: New data and opportunities. In Computer Vision and Pattern Recognition Workshops (CVPRW), 2016 IEEE Conference on 2016 Jun 26 (pp. 1014-1022). IEEE.
\bibitem{Smartereum} Smartereum. \url{https://smartereum.com/}.  Accessed: 2018-08-14.
\bibitem{Stark} Stark, J.; Broomhead, D.S.; Davies, M.E.; Huke, J. (2003). Delay embeddings for forced systems. II. Stochastic forcing. Journal of Nonlinear Science, 13(6), 519--577.
\bibitem{Takens1981} Takens, F. (1981). {Detecting strange attractors in turbulence}.
In {\em Dynamical systems and turbulence, Warwick 1980, Volume 898}, Lecture Notes in Mathematics, Volume 898,  Springer-Verlag, 1981, 366--381.
\bibitem{Truong} Truong, P. (2017). An exploration of topological properties of high-frequency one-dimensional financial time series data using TDA.  KTH Royal Institute of Technology.

\end{thebibliography}
\end{document}